\documentclass{article}
\pdfoutput=1
\usepackage[labelfont=bf]{caption}
\usepackage[text={6.5in,9in}]{geometry}

\usepackage{color}
\usepackage{soul}
\usepackage{pgf}
\usepackage{caption}
\usepackage{subcaption}
\usepackage{wrapfig}
\usepackage{tikz}
\usepackage{amsmath}
\usepackage{url}
\usepackage[square,sort,comma,numbers]{natbib}
\usepackage{comment}
\usepackage{booktabs}
\usepackage{enumitem}
\usepackage{multirow}

\newcommand\blfootnote[1]{
  \begingroup
  \renewcommand\thefootnote{}\footnote{#1}%
  \addtocounter{footnote}{-1}%
  \endgroup
}
\newcommand{\lbin}[0]{\textit{left}}
\newcommand{\clbin}[0]{\textit{center-left}}
\newcommand{\cbin}[0]{\textit{center}}
\newcommand{\crbin}[0]{\textit{center-right}}
\newcommand{\rbin}[0]{\textit{right}}
\newcommand{\ToolX}[0]{CenterTube}
\newcommand{\numsockpuppets}{100,000}
\newcommand{\numtrainingvideos}{9,930,110}
\newcommand{\numtestingvideos}{5,393,820}
\newcommand{\numtotalvideos}{15,323,930}
\newcommand{\numuniqtrainingvideos}{23,735}
\newcommand{\numuniqtestingvideos}{381,153}
\newcommand{\numuniqtotalvideos}{399,935}
\newcommand{\numtrainingchannels}{9,848,784}
\newcommand{\numtestingchannels}{5,279,996}

\newcommand{\numuniqtrainingchannels}{1,256}
\newcommand{\numuniqtestingchannels}{111,450}
\newcommand{\numuniqtotalchannels}{111,715}
\newcommand{\numtrainingslants}{23,715}
\newcommand{\numtestingslants}{125,341}

\newcommand{\trainingcoverage}{99.92\%}
\newcommand{\testingcoverage}{32.88\%}

\newcommand{\uncoveredslantnotweets}{26.79\%}
\newcommand{\uncoveredslantnotenoughlandmarks}{19.47\%}
\newcommand{\uncoveredslantnolandmarks}{17.71\%}

\newcommand{\kstestmoderateseedfarleft}{0.025*}
\newcommand{\kstestmoderateseedleft}{0.014*}
\newcommand{\kstestmoderateseedright}{0.020*}
\newcommand{\kstestmoderateseedfarright}{0.060*}
\newcommand{\kstestsameseedfarleft}{0.127*}
\newcommand{\kstestsameseedleft}{0.109*}
\newcommand{\kstestsameseedright}{0.140*}
\newcommand{\kstestsameseedfarright}{0.228*}
\newcommand{\homepagefarleftmedian}{-0.25}
\newcommand{\homepageleftmedian}{-0.14}
\newcommand{\homepagemoderatemedian}{-0.04}
\newcommand{\homepagerightmedian}{0.09}
\newcommand{\homepagefarrightmedian}{0.29}
\newcommand{\kstesthomepagefarleft}{0.144*}
\newcommand{\kstesthomepageleft}{0.078*}
\newcommand{\kstesthomepageright}{0.109*}
\newcommand{\kstesthomepagefarright}{0.206*}
\newcommand{\leftexposurefarleft}{1.32$\times$}
\newcommand{\rightexposurefarleft}{0.95$\times$}
\newcommand{\leftexposureleft}{1.35$\times$}
\newcommand{\rightexposureleft}{0.97$\times$}
\newcommand{\leftexposuremoderate}{1.28$\times$}
\newcommand{\rightexposuremoderate}{0.99$\times$}
\newcommand{\leftexposureright}{1.20$\times$}
\newcommand{\rightexposureright}{1.16$\times$}
\newcommand{\leftexposurefarright}{1.09$\times$}
\newcommand{\rightexposurefarright}{1.37$\times$}
\newcommand{\spearmansrchannels}{0.95}

\begin{document}

\title{YouTube, The Great Radicalizer? \\Auditing and Mitigating Ideological Biases \\in YouTube Recommendations\blfootnote{Additional information, including the accompanying source code and data, is available at: \\\protect\url{https://github.com/ucdavis-noyce/YouTube-Auditing-Mitigating-Bias}}
}

\author{
Muhammad Haroon, Anshuman Chhabra, \\Xin Liu, Prasant Mohapatra, Zubair Shafiq, Magdalena Wojcieszak\\\\
University of California, Davis}

\date{}

\maketitle

\begin{abstract}
      Recommendations algorithms of social media platforms are often criticized for placing users in ``rabbit holes'' of (increasingly) ideologically biased content. Despite these concerns, prior evidence on this algorithmic radicalization is inconsistent. Furthermore, prior work lacks systematic interventions that reduce the potential ideological bias in recommendation algorithms.
      We conduct a systematic audit of YouTube's recommendation system using a hundred thousand sock puppets to determine the presence of ideological bias (i.e., are recommendations aligned with users' ideology), its magnitude (i.e., are users recommended an increasing number of videos aligned with their ideology), and radicalization (i.e., are the recommendations progressively more extreme). Furthermore, we design and evaluate a bottom-up intervention to minimize ideological bias in recommendations without relying on cooperation from YouTube. 
      We find that YouTube's recommendations do direct users -- especially right-leaning users -- to ideologically biased and increasingly radical content on both homepages and in up-next recommendations. Our intervention effectively mitigates the observed bias, leading to more recommendations to ideologically neutral, diverse, and dissimilar content, yet debiasing is especially challenging for right-leaning users. Our systematic assessment shows that while YouTube recommendations lead to ideological bias, such bias can be mitigated through our intervention. 
\end{abstract}

\section{Introduction}


American society is more divided than ever before---there is a growing gap between the left and the right on key policies \cite{newport2017partisan}, hostility between partisans is increasing \cite{iyengar2019origins}, and support for political violence and rejection of democratic norms are not uncommon \cite{kalmoe2019lethal}. Although many factors contribute to the growing polarization and radicalization, the rise of online social media has come under increased scrutiny. Many critics observe that social media platforms place users in unique environments characterized by self-curated information flows (\textit{selective exposure} \cite{hart2009feeling}), filtered through one's social network (\textit{homophily} \cite{mcpherson2001birds}), and reinforced by recommendation algorithms (\textit{filter bubble} \cite{pariser2011filter}). This may lead to a feedback loop---a potential vicious cycle of reinforcement. The worry is that this \textit{loop effect} is prevalent on social media platforms and is a major contributing factor to polarization and radicalization\footnote{Naturally, most citizens do not inhabit insular communities online \cite{pablo_barbera_social_2020,eady_how_2019}. The small set that does, however, is socially consequential. Those users tend to be more partisan, more politically active, and have a disproportionate influence on the democratic process \cite{abramowitz2008polarization, barbera2019leads}} \cite{chaney2018algorithmic, stray2021designing}. 

Compared to individual biases and social homophily, recommendation algorithms are the least well understood factor that profoundly influences online exposure and behavior \cite{lobaugh2015navigating}. Recommendation algorithms are designed to optimize user activity and engagement on the platform \cite{milano2020recommender} by personalizing recommendations based on users' past exposures and content viewed or shared by other similar users. 
This allows the user to see personally interesting and relevant content. However, issues arise in the case of political content where these factors are feared to reinforce users' political biases, ultimately leading to radicalization, fomenting civil unrest, and endangering public health \cite{chaney2018algorithmic, stray2021designing}. 

In this context, YouTube in particular is receiving increasing scrutiny from scholars, media, and regulators. It is the most popular social media platform, used by 81\% of the U.S. population and with a steadily growing user base \cite{auxier2021socialmediaPEW}. Almost 70\% of content watched on YouTube is recommended by its algorithm \cite{rodriguez2018ytrecommendationsQZ}, which is proprietary and opaque to users and regulators. There are concerns that the algorithm exposes users to divisive, unverified, extreme, conspiratorial, and otherwise problematic content \cite{Diaz21TrumpAccountSuspended, McNamee21PlatformsInsurrectionWIRED, roose2019makingytradicalNYT}. Accordingly, YouTube's algorithm has been described as a ``long-term addiction machine'' \cite{roose2019makingytradicalNYT} and YouTube is accused of putting its users in ``rabbit holes'' and claimed to be ``one of the most powerful radicalizing instruments of the 21st century'' \cite{tufekci2018youtube}. 

Despite immense potential impact of YouTube recommendation algorithm, two key questions remain understudied. First, to what extent do human biases reflected in a user’s watch history drive ideologically biased recommendations? Evidence as to whether YouTube's algorithm disproportionately recommends ideologically biased content is inconclusive. Whereas some studies report that the algorithm tends to recommend radical, divisive, or conspiratorial content \cite{hussein2020misinformation,sanna2020yttrex,papadamou2020just,papadamou2020over}, others provide evidence to the contrary \cite{ribeiro2020auditing,ledwich2020algorithmic}. These conflicting conclusions are due to subtle but crucial differences in the methodologies. In particular, some work relies on active measurements using untrained sock puppets (i.e., without any watch history) \cite{ribeiro2020auditing, ledwich2020algorithmic, brown2021youtuberecommendationworkingpaperNYU}, and thus cannot capture recommendation processes among actual users. In turn, the studies that measure real user watch activity cannot tease apart the role of algorithmic recommendations from the actions of the user \cite{ribeiro2020auditing, hosseinmardi2021examining, chen2021adlreport}. 

Second, how to design interventions that can effectively reduce bias and radicalization? Extant attempts are mainly top-down interventions designed from the perspective of the recommendation system designers and thus can only be employed by the platforms themselves \cite{steck2010training, marlin2009collaborative, chen2018social, wang2018deconfounded, saito2020asymmetric, sun2019debiasing, fabbri2020effect}. These, however, are not incentivized to adapt top-down interventions that might hurt their bottom line. As such, there is a clear unmet need to design \textit{bottom-up} interventions that can be deployed without relying on cooperation from social media platforms. 

We addresses these gaps and advance past work in two key ways. We design a sock puppet based auditing methodology that allows us to systematically isolate the influence of the algorithm in ideologically biased and radicalized recommendations at scale. 
Trained sock puppets achieve a middle ground between untrained sock puppets and real users and can reconcile conflicting evidence in prior literature on the role of YouTube's algorithm in radicalization. We trained \numsockpuppets{} sock puppets, watching a total of \numtrainingvideos{} YouTube Videos from \numuniqtotalchannels{} channels, to reflect different ideology categories (from \lbin{} to \rbin{}). Our large-scale audit examines ideological bias (i.e., are recommendations aligned with users' ideology), its magnitude (i.e., are users recommended an increasing number of videos aligned with their ideology), and radicalization (i.e., are the recommendations progressively more extreme). We provide empirical evidence of biased recommendations by YouTube's algorithm on each of our outcomes (i.e., ideological bias, its magnitude, and radicalization) for both homepage and up-next recommendations. Some of these biases are especially pronounced for right-leaning users.

Based on the audit, we investigate principled interventions to mitigate bias in YouTube's recommendations. In particular, our intervention monitors the recommendations for a given user for signs of ideological bias and then mitigates that bias by watching additional videos in a principled manner using a reinforcement learning model. Through prior training, this model has learnt which type of video optimally reduces the observed bias in recommendations. The selected videos are then played in the background during a time when the user is not interacting with YouTube for a seamless user experience. We focus on introducing more ideologically neutral, diverse, and/or dissimilar content to users' YouTube experience so that the resulting recommendations are ideologically balanced. Democratic theorists have long argued that encountering diverse array of information and sources is beneficial. It is said to promote ``representative thinking'' \cite{Arendt1968}, ``sound political judgment'' \cite{Page1996}, and ``enlightened understanding'' \cite{Dahl1989}, as well as transform citizens into a cohesive collective\footnote{Establishing clear empirical benchmarks for such diversity is difficult \cite{joris2020news} and what aggregate balance of neutral, congenial, or opposing views is most beneficial remains an open normative question. Nevertheless, there is general agreement that citizens need to be familiar with a range of viewpoints.} \citep{Barber1984}.
We find that our intervention effectively minimizes ideological bias in YouTube's algorithm and prompts more ideologically balanced, on average, content recommendations. This debiasing, however, is more challenging for right-leaning users.

\section{Background \& Related Work}
\label{sec: related work}
In this section, we first survey prior literature on various sources of bias in recommendation systems as well as the landscape of potential interventions. 
We then summarize recent efforts to audit YouTube's recommendation system to uncover its role in perpetuating political bias, radicalization, and misinformation.

\subsection{Background}
\begin{wrapfigure}{R}{0.35\textwidth}
    \centering
    \includegraphics[width=0.35\textwidth]{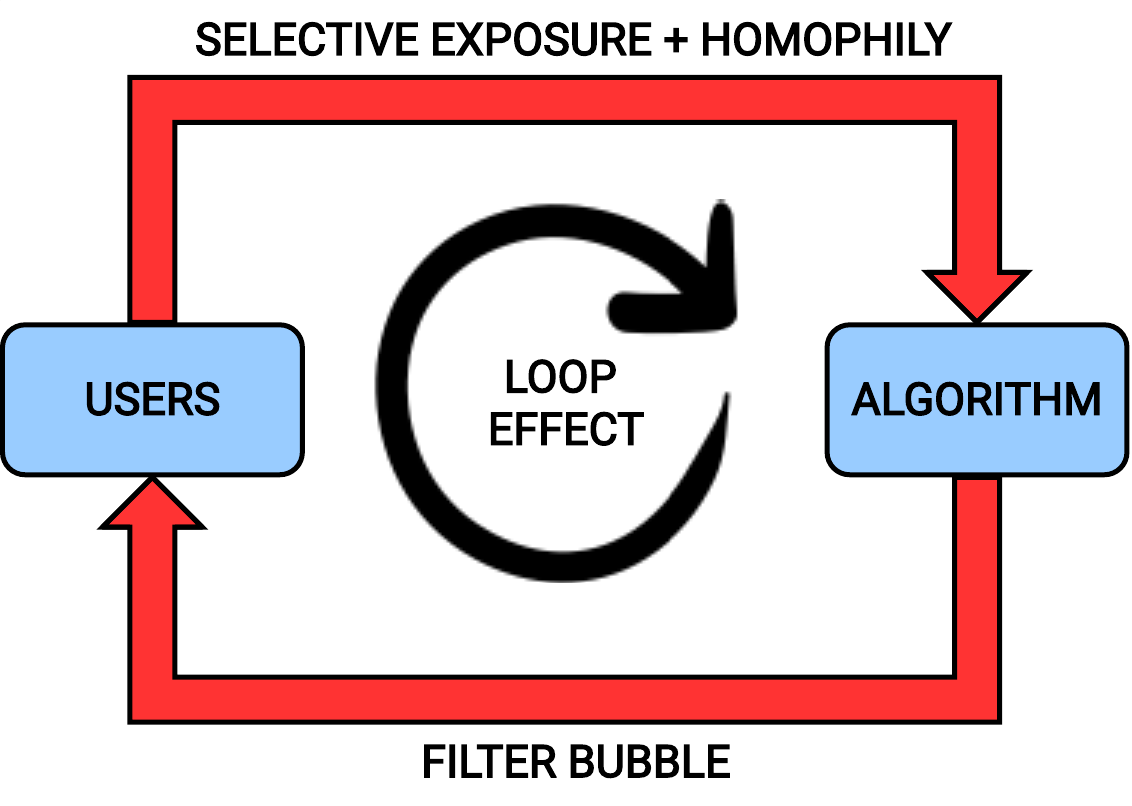}
    \caption{Illustration of the loop effect in social recommendation systems.}
    \label{fig:loop}
\end{wrapfigure}

There are many potential sources of bias in a social recommendation system that can be attributed to individual users (selective exposure \cite{hart2009feeling}), social network (homophily \cite{mcpherson2001birds}), or the recommendation algorithm (filter bubble \cite{pariser2011filter}). 
As shown in Figure \ref{fig:loop}, these factors interact in a closed-loop recommendation system, potentially forming a vicious cycle addressed in this work.


\noindent $\bullet$ \textbf{Selective Exposure:} Users with certain beliefs tend to interact with social media content that is consistent with those beliefs. This is partly because inconsistencies in one's beliefs generate discomfort and a desire to reduce it \cite{festinger1957theory,garrett2009politically}. Although the extent to which people in fact seek out like-minded content is debated \cite{wojcieszak2020social,guess2021almost}, users are more likely to select content that aligns with their preexisting beliefs than otherwise. This tendency may polarize people's attitudes and generate hostility toward political out-groups \cite{levendusky2013partisan1, levendusky2013partisan2}.

\noindent $\bullet$ \textbf{Homophily:} Social networks, both offline and online, are largely curated by an individual. According to the well-documented homophily principle, people's networks and relationships are often homogeneous with regard to socio-demographic, behavioral, and political characteristics \cite{mcpherson2001birds}. On social media, users are more likely to follow and exchange political information with other users of similar ideology \cite{barbera2015tweeting, colleoni2014echo, conover2011political, conover2012partisan, fabbri2020effect, mosleh2021shared}. These homophilous social networks act as an information filter that further reduces the spectrum of content the user encounters \cite{bakshy2015exposure}. Thus, homophily may reinforce preexisting biases from selective exposure. 
While a majority of the U.S. population does not inhabit online echo chambers and does not even engage with politics on social media  and online in general \cite{pablo_barbera_social_2020,eady_how_2019}, a small and potentially more consequential subset of the population does so \cite{Gentzkow_Shapiro_2011, flaxman2016filter, wojcieszak2021no}. 

\noindent $\bullet$ \textbf{Filter Bubble:} Social recommendation algorithms are designed to maximize user activity and engagement on the platform \cite{milano2020recommender}. To this end, they customize each user's experience by personalizing recommendations based on the user's past exposure and engagement as well as content shared by other similar users in their social network. These personalized recommendations influence the content users consume \cite{lobaugh2015navigating} and -- for some users -- may further reinforce users' biases from selective exposure and homophily, leading to the ``filter bubble'' effect \cite{pariser2011filter}. 

\noindent $\bullet$ \textbf{Loop Effect:} Putting it all together, the loop effect represents the vicious cycle, in which social recommendation systems become increasingly ideologically biased over time, i.e., recommending political content that is aligned with users' past exposure on the platform and resulting in ideologically like-minded exposures.  

\subsection{Related Work}

These concerns are particularly relevant to YouTube, one of the most popular online social media platforms. YouTube has been accused of putting its users in ``rabbit holes'' \cite{ledwich2020algorithmic} and its algorithm has been described as a ``long-term addiction machine'' \cite{roose2019makingytradicalNYT}. This is a grave concern because 70\% of watched content on YouTube is via recommendations \cite{rodriguez2018ytrecommendationsQZ} and the top recommendation is typically played automatically after the currently-watched video.
Thus, the fears are, that unknowingly and unwillingly, some audiences encounter more and more ideologically biased content and partisans are at risk of being further radicalized through the actions of the recommendation system. The work on the existence of ideological bias on YouTube is not fully conclusive, as we detail below. We also outline extant attempts to mitigate bias in recommendations and specify how our work extends these attempts.

\subsubsection{Auditing YouTube's recommendation system} 
To systematically examine the alleged ideological biases in algorithmic recommendations and their potential role in user radicalization, there have been several audits conducted by the academic community. The recent audits, however, report mixed evidence as to whether YouTube's recommendation algorithm is to blame for radicalization. This mixed evidence is largely due to methodological differences between these audits.

One line of research studies YouTube through active measurements using sock puppets. For instance, Ribeiro et al. \cite{ribeiro2020auditing} used untrained sock puppets (i.e., without any watch history) to show that about 5\% algorithmic pathways led from ideologically moderate to extreme channels through video and channel recommendations. Similarly, Ledwich et al. \cite{ledwich2020algorithmic} also relied on untrained sock puppets to show that 16\% of the algorithmic pathways (11\% moderate$\rightarrow$left and 5\% moderate$\rightarrow$right) led from ideologically moderate to extreme channels through video recommendations. 
%
%
However, these studies assumed that YouTube's personalized recommendations for trained sock puppets are not different from those for untrained sock puppets \cite{chen2019top, papadamou2020just}. This, we argue, likely underestimates the prevalence of algorithmic radicalization pathways. 

Another line of research has used passive measurements of real user activity. An analysis of video comments by Ribeiro et al. \cite{ribeiro2020auditing} shows that a significant percentage of users migrate from ideologically moderate to more extreme channels on YouTube. 
Hosseinmardi et al. \cite{hosseinmardi2021examining} longitudinally analyzed browsing history of more than 300,000 users to also show that a significant percentage of users migrate from moderate to extreme videos. They also found that 8.5\% and 31.8\% of extreme video views originate from homepage and video recommendations, respectively. Furthermore, in a report for the Anti-Defamation League (ADL), Chen et al. \cite{chen2021adlreport} analyzed browsing history of 915 users to show that 9.2\% of users watched a video from an extreme channel but less than 1\% of the recommendations that users actually followed were from an extreme channel. 
This line of work is very important. However, by using real user's browsing histories, it cannot tease apart the role of algorithmic recommendations from the actions of the user in exposure to problematic content on YouTube. 


On one hand, audits analyzing sock puppets disregard the impact of sock puppet training, which is crucial for personalized recommendations and observing the loop effect in practice.
%
On the other hand, audits analyzing real user activity cannot tease apart the dependencies between algorithmic and others factors driving radicalization. 
As we discuss below, trained sock puppets achieve a middle ground between untrained sock puppets and real users for auditing purposes because (i) they trigger the feedback loop effect that real users (but not untrained sock puppets) experience and (ii) the observed radicalization can be attributed solely to YouTube's recommendations (which is not possible for studies of real users). In fact, recent work has adjusted this framework of trained sock puppets to audit misinformation in YouTube's recommendation system \cite{hussein2020misinformation, papadamou2020just, papadamou2020over, sanna2020yttrex, faddoul2020longitudinal}. We extend this work to the broader consequential domain of ideologically biased recommendations.

\subsubsection{Mitigating bias in YouTube's recommendations}
Given the general concerns from journalists, academics, and policymakers, there have been various attempts to address biases in recommendation algorithms. YouTube has tried to implement interventions \cite{buntain2021ytrecommendations,faddoul2020longitudinal,youtube4rs}. Some evidence suggests, however, that these interventions have not significantly mitigated the problem \cite{buntain2021ytrecommendations, faddoul2020longitudinal}. A likely exception is the recent intervention to mitigate the promotion of COVID-19 vaccine misinformation content. YouTube audits show that prevalence of misinformation recommendations on this topic is less than other topics \cite{papadamou2020just, sanna2020yttrex, hussein2020misinformation}. We note, however, that there is an inherent conflict between the advertising-powered business model of online social media platforms that is geared towards user engagement, which is typically maximized by dialing up biases in recommendation algorithm. In fact, a former YouTube employee revealed that YouTube's recommendation algorithm was actively promoting problematic content because it was more engaging, and in turn helped increase advertising revenues, but internal attempts to mitigate this issue were rebuked \cite{lewis2018exgoogleemployeeguardian,harnett2018exgoogleemployeeKQED}.
%

Prior research on mitigating biased recommendations is limited to top-down interventions, which are designed from the perspective of the recommendation system designers. More specifically, prior work attempts to modify the recommendation algorithm (i.e., in-processing; e.g., architecture, optimization function) \cite{zafar2017fairness, chen2018stabilizing, zhao2018recommendations, burke2018balanced, farnadi2018fairness}, the inputs to the algorithm (i.e., pre-processing; e.g., training dataset) \cite{yu2020influence, liu2020general, rastegarpanah2019fighting, gordaliza2019obtaining}, or the output of the algorithm (i.e., post-processing; e.g., recommendation confidence) \cite{nandy2020achieving, kamiran2012decision, hardt2016equality, pleiss2017fairness} to debias recommendation systems. However, such mitigations can only be employed by the social media platforms themselves, which as discussed above, are not incentivized to do so \cite{steck2010training, marlin2009collaborative, chen2018social, wang2018deconfounded, saito2020asymmetric, sun2019debiasing, fabbri2020effect}.

To our best knowledge, prior work lacks debiasing interventions that can be employed by individual users without requiring any cooperation from the social media platforms. 
Such interventions are particularly challenging to design because they need to operate with only ``blackbox access'' to a production social recommendation system. The research community has only recently started developing such interventions. For instance, offering promising evidence, Tomlein et al. \cite{tomlein2021audit} recently demonstrated that it was possible to ``burst the bubble'' of misinformation videos on YouTube by watching videos debunking that misinformation. Also, Zhang et al. \cite{zhang2022harpo} recently demonstrated that it is possible to mislead user profiling and ad targeting models in online behavioral advertising.

\section{Auditing YouTube Recommendations}
\label{sec:audit}

In order to measure the extent to which YouTube's recommendations are susceptible to ideological biases, we conduct a systematic audit of the platform using ``sock puppets''.
A sock puppet is an automated browser instance that mimics a YouTube user by watching videos and gathering recommendations.
From the results of our audit, we seek to answer the following research questions.
\begin{enumerate}[label=RQ \arabic*:, left=\parindent]
    \item Are recommendations ideologically biased?
    \item Does following the recommendation trail increase exposure to ideologically biased content?
    \item Does following the recommendation trail lead to increasingly radical videos?
\end{enumerate}
We now provide an overview of how we create and operationalize sock puppets to audit YouTube's recommendations in $\S$\ref{sec:overview-audit}.

\subsection{Overview of audit}
\label{sec:overview-audit}

\begin{figure*}[!t]
    \includegraphics[width=\textwidth]{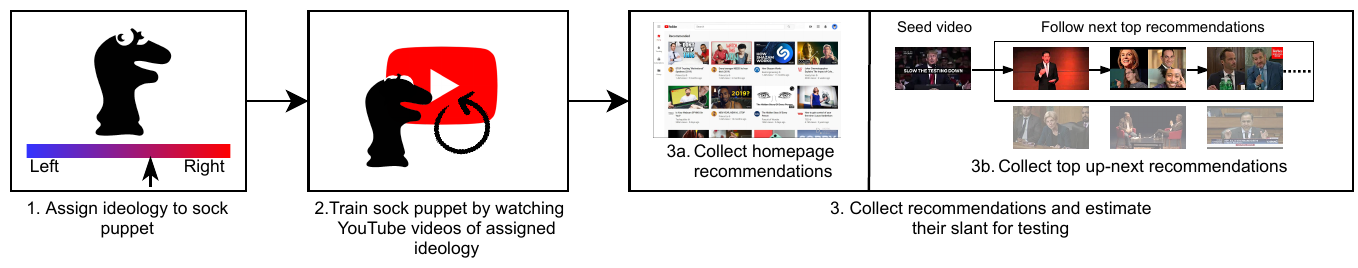}
    \caption{Overview of our sock puppet audit of YouTube's recommendations.}
    \label{fig:overview-audit}
\end{figure*}

An overview of our audit can be seen in Figure \ref{fig:overview-audit}.
The audit consists of two main phases: \textit{training} and then \textit{testing} sock puppets.
Each individual sock puppet is implemented as an automated web browser instance that runs YouTube in an isolated container\footnote{The sock puppets are not logged into a Google account for two reasons:
First, it is infeasible to create a new account for each sock puppet at scale.
Second, YouTube is capable of personalizing recommendations without an account via cookies \cite{googlecookiepolicy}.
}.
The sock puppets are trained by watching videos of a particular political ideology and then their personalized recommendations are tested in a controlled manner. 
Specifically, we test two types of YouTube recommendations \cite{howytrecommendations}: (1) the recommendations on the homepage; and (2) the trail of up-next recommendations starting from a seed video.
We now provide a detailed description of the training and testing phases in $\S$\ref{sec:training-sock-puppets} and $\S$\ref{sec:testing-sock-puppets} respectively, before describing how we estimate the ideology of a YouTube video in $\S$\ref{sec:estimating-video-ideology}.

\subsubsection{Training sock puppets}
\label{sec:training-sock-puppets}
The sequence of videos watched during the training phase reinforce hypothetical pre-existing ideological biases of a user. As we discussed in $\S$\ref{sec: related work}, the existence of these prior ideological biases is assumed by most work as starting the loop effect, and this personalization is key to observing ideological bias and radicalization in algorithmic recommendations.
We train the sock puppets on one of the following five ideologies: \lbin{}, \clbin{}, \cbin{}, \crbin{}, and \rbin{}.
Combined, we train over \numsockpuppets{} sock puppets and each watches a 100 randomly sampled videos from its assigned ideology for 30 seconds each, in accordance with the insight from \cite{papadamou2020just}\footnote{For ads that appear during watch, we skip them if possible and wait until they are over if not.}.
The videos in each ideology are collected from known political and news channels and their ideology is estimated using the approach discussed in $\S$\ref{sec:estimating-video-ideology}.

\subsubsection{Testing sock puppets}
\label{sec:testing-sock-puppets}
Once a sock puppet has been trained, we first collect the video recommendations on the YouTube homepage of the sock puppet.
The homepage is expected to contain video recommendations that are of interest to the sock puppet\footnote{YouTube also uses Google account activity to personalize recommendations \cite{googlemanageyt} but since the sock puppet is not logged in, it is reasonable to assume that the YouTube watch history is the sole factor.}.
Second, we gather the up-next video recommendations from a fixed YouTube video (henceforth, referred to as the \textit{seed video}) \cite{howytrecommendations} and build a \textit{recommendation trail}, mimicking YouTube's auto-play functionality \cite{howytautoplay}.
This recommendation trail represents the algorithmic ``rabbit hole''  \cite{tufekci2018youtube} that is said to lead users to more extreme content.
Because the sock puppet chooses to follow the recommendations, it can get stuck in the vicious cycle of ideologically biased video recommendations --- thus, forming the basis of the loop effect.
\subsubsection{Estimating video ideology}
\label{sec:estimating-video-ideology}
We estimate the ideological slant of a YouTube video by analyzing its audience on Twitter, an adaptation of the approach by Le et al. \cite{le2017scalable}. Specifically, we collect the tweets mentioning a given YouTube video using the Twitter API \cite{twitter2021api} and check if the authors of those tweets follow a set of well-recognized partisan elites, i.e., landmarks, whose political ideology is clearly established as either liberal or conservative.\footnote{
The list of landmarks is also obtained from Le et al. \cite{le2017scalable} and, following their recommendations, we only estimate the slant if the number of landmarks is greater than 12.}
The landmarks followed by a user provide insight into the user's political ideology as Twitter users are more likely to follow other users of the same ideology \cite{barbera2015birds}. 
For the tweets mentioning a given YouTube video, we count the total number of liberal ($L$) and conservative ($C$) landmarks followed by the authors of those tweets and estimate the slant on a scale of -1 (most liberal) to +1 (most conservative) as:
\begin{equation}
Slant = \dfrac{C - L}{C + L}
\label{equation: slant}
\end{equation}
In addition to their continuous slant estimate, we also categorize videos in the following slant ranges $[-1, -0.6)$, $[-0.6, -0.2)$, $[-0.2, +0.2)$, $[+0.2, +0.6)$, $[+0.6, +1]$ as \lbin{}, \clbin{}, \cbin{}, \crbin{}, and \rbin{} respectively corresponding to the sock puppet ideologies specified during training.

\begin{figure}
    \begin{minipage}[t]{0.475\textwidth}
        \centering
        \includegraphics[width=\textwidth]{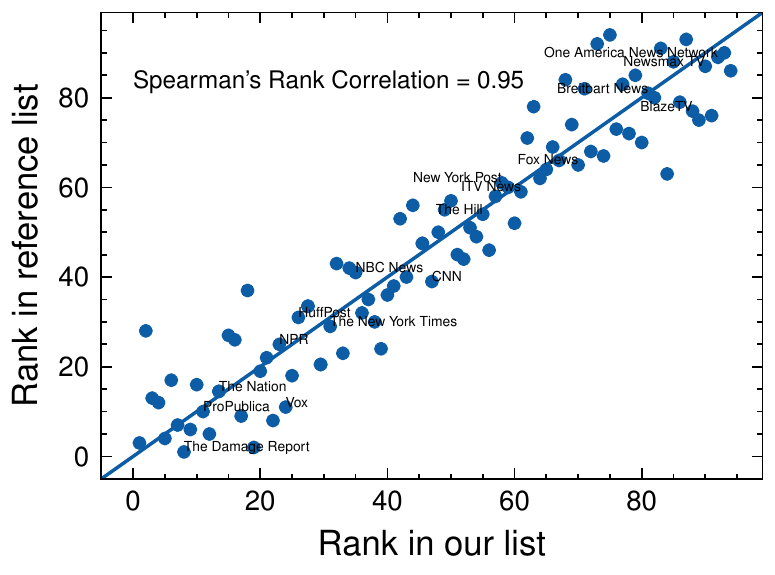}
        \caption{Rank correlation of YouTube channels and Twitter accounts from the reference list.}
        \label{fig:spearmansr-correlation}
    \end{minipage}
    \hfill
    \begin{minipage}[t]{0.475\textwidth}
        \centering
        \includegraphics[width=\textwidth]{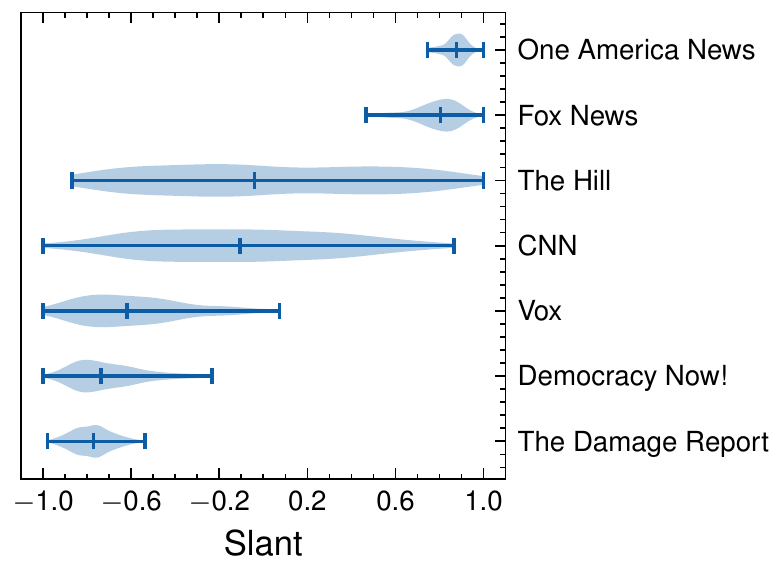}
        \caption{Violin plot for the distribution of slant estimations for the top-100 most popular videos in the channel. The density of the plot at a given slant indicates the number of videos in the distribution around that point. The middle tick represents the mean of the distribution.}
        \label{fig:video-slant-distribution}
    \end{minipage}
\end{figure}

We assess the validity of our video-level slant estimates by comparing them with the Twitter-based estimates provided by McCabe et al.\footnote{\url{https://github.com/sdmccabe/new-tweetscores/releases/tag/v0.1.0}} based on the method developed by Barberá et al. \cite{barbera2015tweeting}.
For the Twitter accounts in their list, we first identify the corresponding YouTube channel and then compute the mean slant of the top-100 most popular videos of that channel.\footnote{We exclude channels with less than a 100 videos from this comparison due to lack of sufficient data.}
%
%
Ideally, when ordered from most liberal to most conservative, the rank of the Twitter accounts in the baseline list should be the same as the rank of YouTube channels in ours.
We plot the relative ranks by both methods in Figure \ref{fig:spearmansr-correlation}.
It is evident that the ranks follow closely along with the diagonal ($x = y$) indicating a high correlation between the ranks of the YouTube channels based on our slant estimation method and the Twitter accounts.
We quantify this correlation by computing Spearman's rank order metric \cite{spearmansr2008}, which at $\spearmansrchannels{}$ indicates that the ranks of two lists ordered liberal-to-conservative strongly co-vary.
These results demonstrate the validity of our YouTube channel-level slant estimation against the state-of-the-art \cite{robertson2018auditing}.

It is important to emphasize that our approach substantially extends prior work, which analyzes political ideology of YouTube videos at a coarse channel-level granularity  \cite{ribeiro2020auditing, ledwich2020algorithmic, hosseinmardi2021examining}. Our method is able to estimate slant at the finer granularity of individual YouTube videos. To illustrate the importance of video-level slant estimation, Figure \ref{fig:video-slant-distribution} plots the distribution of the slants estimated for the most popular videos from six well-known channels.
Note that while the distributions for the most ideologically extreme channels (i.e., One America News Network (OANN) and Democracy Now!) are skewed and relatively concentrated, the distributions of ideologically moderate channels (e.g., The Hill and CNN) have a large spread. This suggests that simply assigning the overall slant of a YouTube channel to all of its videos can lead to inaccurate estimates. 
This limitation applies to several prior audits which, for example, labelled all videos from CNN as \textit{left} \cite{hosseinmardi2021examining, ribeiro2020auditing,ledwich2020algorithmic}. In contrast, our method clearly shows that not all videos from CNN are ideologically left. 
For example, we observed several instances of anti-Obama/pro-Trump videos from CNN that were correctly labeled as right by our slant estimation method. 
Thus, we conclude that channel-level slant estimation does not accurately represent the ideology of a wide range of videos posted by the channel.\footnote{Green et al. \cite{greencuration} had a similar finding for estimating the slant of newspaper articles versus news domains from which these articles come.}

\subsection{Data collection}

\begin{wraptable}{r}{0.56\textwidth}
\vspace{-.2in}
  \caption{Data collection statistics for \numsockpuppets{} sock puppets}
  \label{tab:statistics}
  \begin{tabular}{|llll|}
    \hline
    & Training & Testing & Total \\
    \hline
  \# of videos & \numtrainingvideos{} & \numtestingvideos{} & \numtotalvideos{} \\
  \# of unique videos & \numuniqtrainingvideos{} & \numuniqtestingvideos{} & \numuniqtotalvideos{} \\
  \# of channels & \numuniqtrainingchannels{} & \numuniqtestingchannels{} & \numuniqtotalchannels{} \\
     \hline
    \end{tabular}
\end{wraptable}

Table \ref{tab:statistics} provides an overview of the data collected from our sock puppets during the training and testing phases of the audit.
We train and test a total of \numsockpuppets{} sock puppets.
In training, each sock puppet is assigned an ideology (i.e., \lbin, \clbin, \cbin, \crbin, and \rbin) and watches 100 randomly sampled videos of its assigned ideology.
The videos used to train sock puppets are from a curated list of political channels  \cite{robertson2018auditing} and are assigned ideology based on their estimated slant.
Our sock puppets encounter a total of \numtrainingvideos{} videos (\numuniqtrainingvideos{} unique) from \numtrainingchannels{} channels (\numuniqtrainingchannels{} unique) in the training phase.

In testing, we use the trained sock puppets to gather both homepage and up-next recommendations.
To collect up-next recommendation trails, we use a curated list of over 200 seed videos containing 40 videos each from the 5 ideological categories.
%
%
We uniformly randomly select a seed video from a predetermined list and then collect the recommendation trail for a depth of up to 20 videos.
Our sock puppets encounter \numtestingvideos{} videos (\numuniqtestingvideos{} unique) from the \numtestingchannels{} channels (\numuniqtestingchannels{} unique) in the testing phase.

In summary, we collect a total of \numtotalvideos{} videos (\numuniqtotalvideos{} unique) spanning \numuniqtotalchannels{} unique channels from \numsockpuppets{} sock puppets.

\subsection{Evaluation}

\subsubsection*{RQ 1: Are recommendations ideologically biased?}

\begin{figure*}[!t]
    \centering
    \begin{subfigure}[b]{0.32\textwidth}
        \centering
        \includegraphics[width=\textwidth]{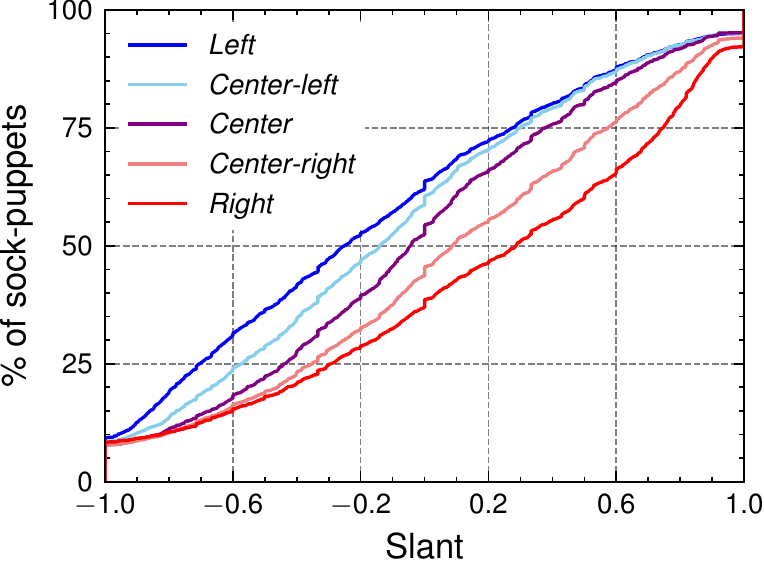}
        \caption{Homepage recommendations\break}
        \label{fig:homepage-cdf}    
    \end{subfigure}
    \hfill
    \begin{subfigure}[b]{0.32\textwidth}
        \centering
        \includegraphics[width=\textwidth]{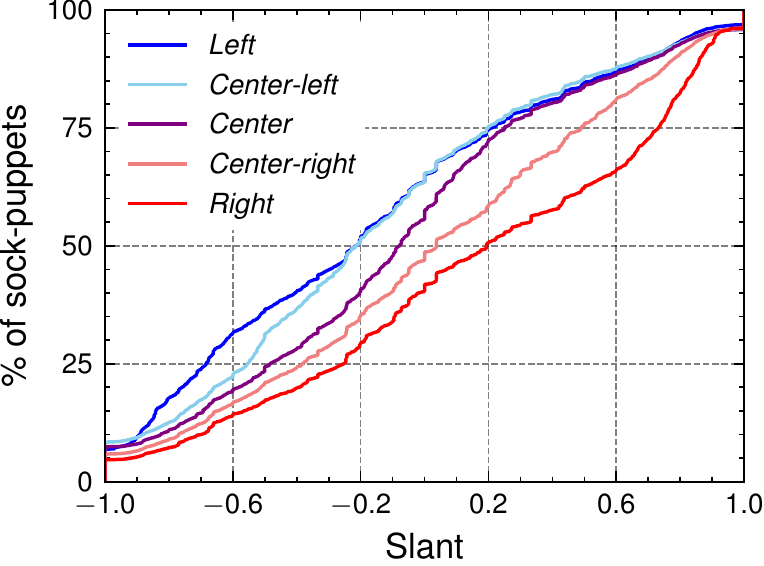}
        \caption{Up-next recommendations with same seed ideology}
        \label{fig:watched-cdf}
    \end{subfigure}
    \hfill
    \begin{subfigure}[b]{0.32\textwidth}
        \centering
        \includegraphics[width=\textwidth]{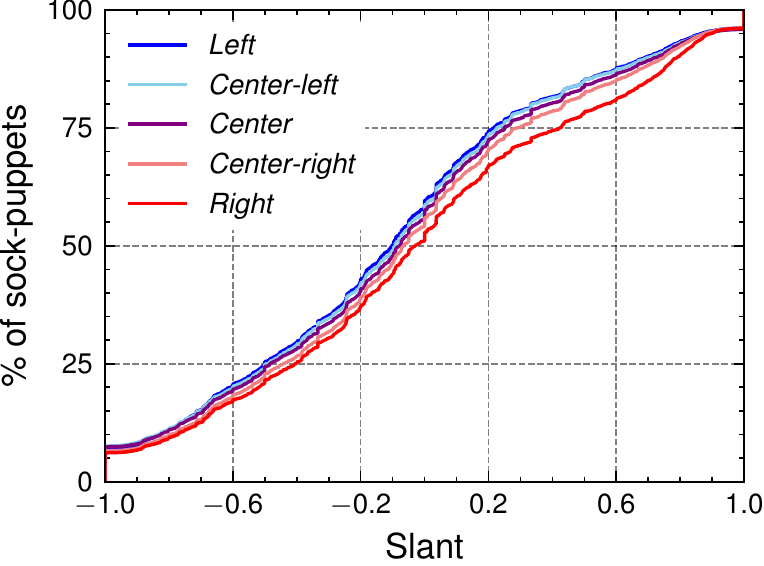}
        \caption{Up-next recommendations with \cbin{} seed ideology}
        \label{fig:watched-cdf-moderate}
    \end{subfigure}
    \caption{Cumulative Distribution Function (CDF) for the videos collected from the sock puppets. The x-axis shows the slant range from [$-1,+1$] and y-axis shows the percentage of videos that have $slant \leq x$.}
    \label{fig:cdf}
    \vspace{-.2in}
\end{figure*}

Ideological bias in recommendations is indicated by the presence of a large number of video recommendations belonging to that ideology.
To answer the first question regarding the presence of ideological bias, we look at the distributions of the slant of the recommended videos for the 5 different types of sock puppets.
Not only do we consider both types of video recommendations: homepage and up-next, we also split up-next recommendations into two types based on the ideology of the seed video.

\vspace{.05in}
\noindent
\textbf{Homepage Recommendations.}
We plot the distribution of slant estimates for the homepage recommendations of each sock puppet ideology in Figure \ref{fig:homepage-cdf}. 
There is a clear visual difference between the slant distributions of the different categories of sock puppets.
From left-to-right, the medians for \lbin{}, \clbin{}, \cbin{}, \crbin{}, and \rbin{} sock puppets are \homepagefarleftmedian{}, \homepageleftmedian, \homepagemoderatemedian{}, \homepagerightmedian{}, and \homepagefarrightmedian{} respectively.
This shows that homepage recommendations are generally biased towards the sock puppet's ideology.

Despite being slightly left, the median of the \cbin{} slant distribution is close to zero (\homepagemoderatemedian{}) and can be used as a baseline to measure the ideological bias of the other sock puppet ideologies.
To assess the statistical significance of differences between these distributions, we use a two-sided pairwise Kolmogorov-Smirnov test between the \cbin{} sock puppet and each of the other ideologies.
The KS-test statistic measures the maximum difference between two cumulative distributions and computes a $p$-value based on their sample sizes.
%
%
If $p$ value is small, say $< 0.05$, then we reject the null hypothesis that the home recommendations are drawn from the same unbiased distribution as the \cbin{} sock puppet.

\begin{wraptable}{r}{0.53\textwidth}
\caption{Statistic values for the Kolmogorov-Smirnov test between the \cbin{} and other sock puppet ideologies. Asterisks (*) indicate significant difference between \cbin{} and the corresponding sock puppet ($p < 0.05$).}
\begin{tabular}{|l|ccc|}
    \hline
    {Ideology} & \small{Homepage} & \multicolumn{2}{c|}{\small{Up-next}}\\
    {} & & \small{Same seed} & \small{\cbin{} seed}\\
    \hline
    \small{\lbin{}}  &   \kstesthomepagefarleft{} & \kstestsameseedfarleft{}& \kstestmoderateseedfarleft{} \\
    \small{\clbin{}} &   \kstesthomepageleft{} &  \kstestsameseedleft{}&  \kstestmoderateseedleft{} \\
    \small{\crbin{}} &   \kstesthomepageright{} &  \kstestsameseedright{}&  \kstestmoderateseedright{} \\
    \small{\rbin{}}  &   \kstesthomepagefarright{} &  \kstestsameseedfarright{}&  \kstestmoderateseedfarright{} \\
    \hline
\end{tabular}    
\label{tab:kstest-audit}
\end{wraptable}
Table \ref{tab:kstest-audit} shows that $p < 0.05$ for all the pairwise tests between the \cbin{} sock puppet and other categories. Thus, we conclude that the ideological bias of homepage recommendations for  \lbin{}, \clbin{},  \crbin{}, and \rbin{} sock puppets are significantly different than that of \cbin{} sock puppets. 
The test statistic further indicates that the magnitude of difference is much larger for \rbin{} sock puppets as compared to \lbin{} sock puppets. 

%
%
We note a visual gap between the \lbin{} and \clbin{} distributions at the $-0.6$ slant, the cut-off between the two categories, suggesting that the \lbin{} received more left-biased recommendations than the \clbin{} sock puppet.
This gap decreases around $-0.4$ where the \clbin{} sock puppet starts seeing similar recommendations before eventually overlapping with the \lbin{} sock puppet around $0$.
Overall, the higher-value for the \lbin{} sock puppet at $-0.8$ shows that it received more left-biased recommendations than the other sock puppets.

%
%
%
%
In comparison, the divergence between the \crbin{} and the \rbin{} sock puppets is higher suggesting that the \rbin{} sock puppet receives more right-leaning recommendations than the \crbin{} sock puppet.
The statistic value for the \rbin{} sock puppet is not only significant but is also greater than that of the other sock puppets.
This indicates that the recommendations for the \rbin{} sock puppet are not only significantly different from the \cbin{} baseline, but are even more ideologically biased than the recommendations of the other sock puppets.

\vspace{0.05in} \noindent \textbf{Up-next recommendations.} 
We turn to analyzing the ideological bias of up-next recommendations in Figures \ref{fig:watched-cdf} and \ref{fig:watched-cdf-moderate}.
As before, we plot the distributions of slant estimates for up-next recommendations for all the sock puppets separately for each ideology category.
This is accomplished by aggregating the top up-next recommendation at each step of the trail into a single distribution for each sock puppet.
Figure \ref{fig:watched-cdf} shows these distributions of up-next recommendations for trails starting from a seed video with the same ideology as the sock puppet's ideology.
This represents a scenario where the sock puppet continues to watch videos of the same ideology as the training but now follows the recommendation trail instead.
We observe increased ideological bias in the up-next recommendation trails compared to the homepage provided that the seed video is of the same ideology.
The fact that the seed video shares the same ideology as the sock puppet is consistent with the general fear and some observations that some partisan YouTube users select, watch, search for, and/or are recommended videos that are politically like-minded. However, an argument can be made that the seed video itself is either responsible for or reinforces the bias in ideology that is present in Figure \ref{fig:watched-cdf}.

To this end, we analyze how the recommendation trails vary if the seed video belongs to the \cbin{} category instead.
Figure \ref{fig:watched-cdf-moderate} shows the cumulative distribution function (CDF) for the up-next recommendations of each sock puppet category starting from a video belonging to the \cbin{} category. Although there is a drop in the ideological bias of the up-next recommendations, the bias is still visible and computing the KS-test statistic between the \cbin{} and the other sock puppets in Table \ref{tab:kstest-audit}, we see that the set of up-next recommendations for different sock puppets is different from the \cbin{} sock puppet and this difference is statistically significant. Furthermore, we note that the statistic value for the \rbin{} sock puppet is again higher than for the other sock puppets, indicating that the increasing number of ideologically biased recommendations is again higher for the \rbin{} sock puppet than it for the others.

\subsubsection*{RQ 2: Does following the recommendation trail increase exposure to ideologically biased content?}

We analyze the role of depth by defining radicalization at a certain depth as the ratio between the extreme recommendations at that depth and at the start of the trail.
In essence, this captures how much more biased content the sock puppets encountered provided that the seed videos were uniformly randomly sampled from each of the 5 ideology categories.
We compute the following metric over the aggregate seeds and up-next recommendations collected for each sock puppet category.

\begin{equation}\label{exposure_eq}
    E^{ideology}_{depth} = \frac{V^{ideology}_{depth}}{V^{ideology}_0}
\end{equation}
    
Here, $E$ refers to exposure and $V$ refers to the distrbution of video recommendations of a specific ideology.
The parameter $ideology \in \{\lbin{}, \rbin{}\}$ defines which ideology is the user radicalized to whereas the parameter $depth$  defines the depth of the trail and $depth=0$ refers to the seed video.
Note that the $depth$ parameter has no upper bound as users can choose to continue following the recommendation trail for as long as they wish.
However, as mentioned earlier, we only follow the trail to a depth of 20 recommendations.

\begin{figure*}[h]
    \centering
    \begin{subfigure}[b]{0.8\textwidth}
        \centering
        \includegraphics[width=\textwidth]{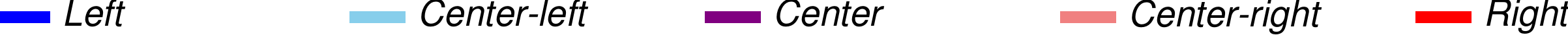}
    \end{subfigure}

    \begin{subfigure}[b]{0.18\textwidth}
        \centering
        \includegraphics[width=\textwidth]{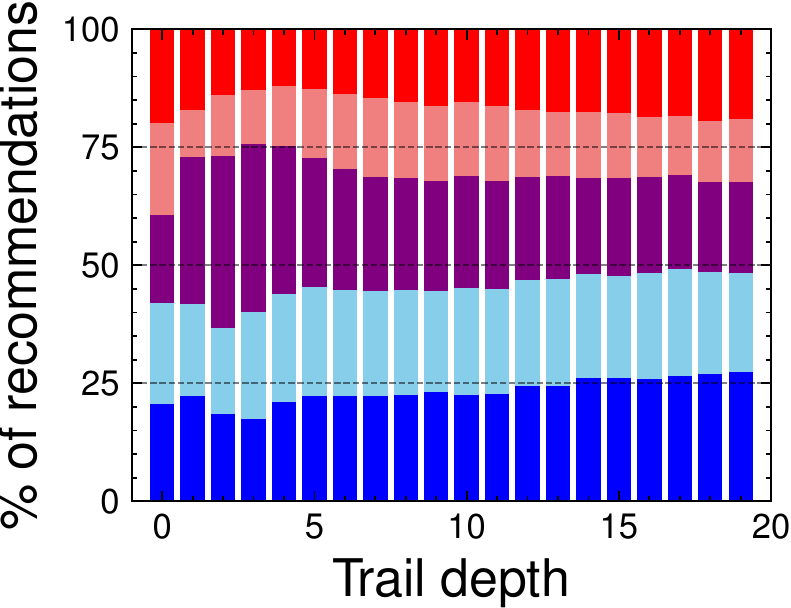}
        \caption{\lbin{} \\ $E^{\lbin{}}_{20} = \text{\leftexposurefarleft{}}$\\ $E^{\rbin{}}_{20} = \text{\rightexposurefarleft{}}$ }
    \end{subfigure}
    \hfill
    \begin{subfigure}[b]{0.18\textwidth}
        \centering
        \includegraphics[width=\textwidth]{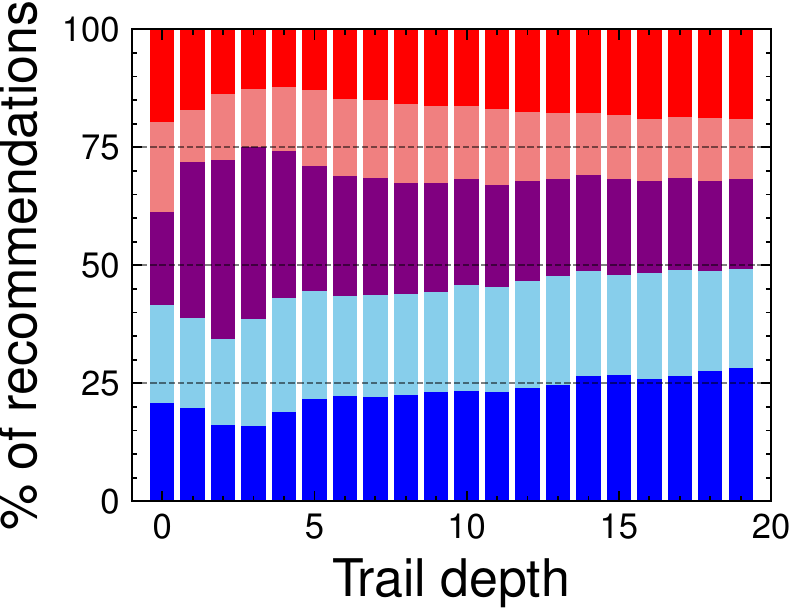}
        \caption{\clbin{} \\ $E^{\lbin{}}_{20} = \text{\leftexposureleft{}}$\\ $E^{\rbin{}}_{20} = \text{\rightexposureleft{}}$ }
    \end{subfigure}
    \hfill
    \begin{subfigure}[b]{0.18\textwidth}
        \centering
        \includegraphics[width=\textwidth]{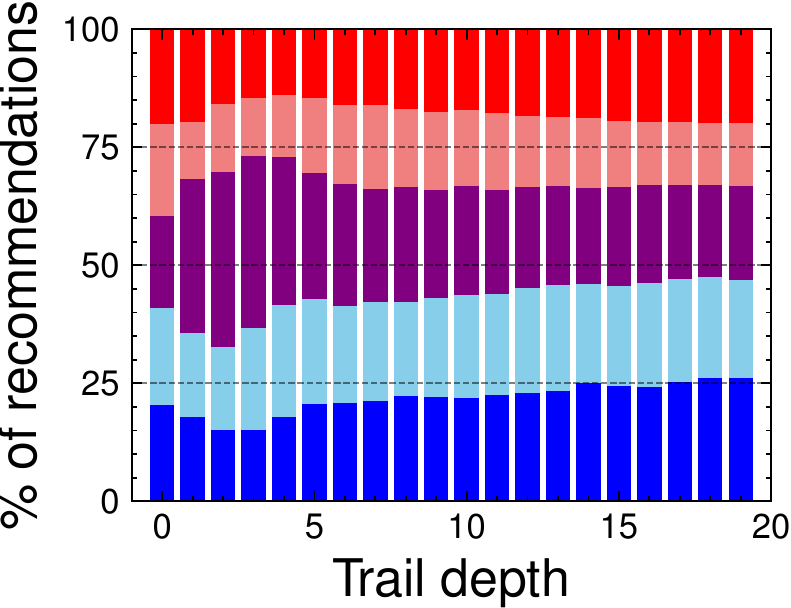}
        \caption{\cbin{} \\ $E^{\lbin{}}_{20} = \text{\leftexposuremoderate{}}$\\ $E^{\rbin{}}_{20} = \text{\rightexposuremoderate{}}$ }
    \end{subfigure}
    \hfill
    \begin{subfigure}[b]{0.18\textwidth}
        \centering
        \includegraphics[width=\textwidth]{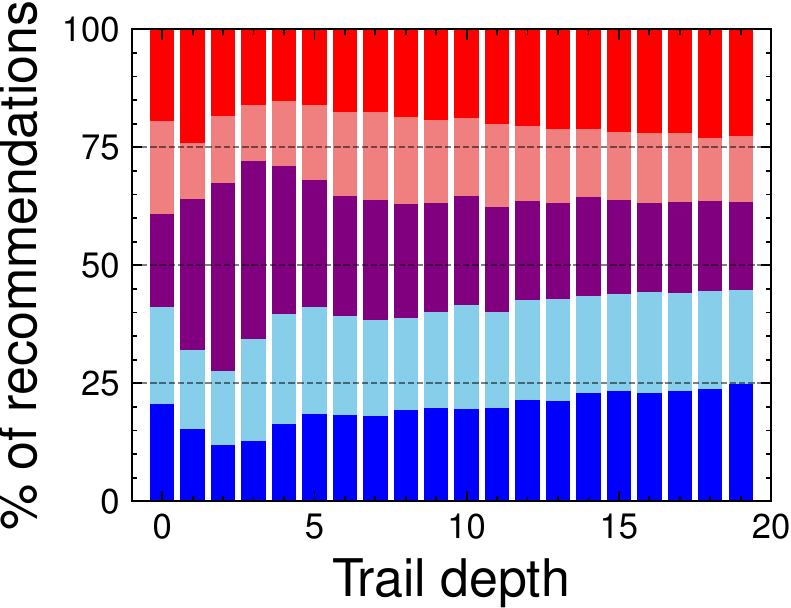}
        \caption{\crbin{} \\ $E^{\lbin{}}_{20} = \text{\leftexposureright{}}$\\ $E^{\rbin{}}_{20} = \text{\rightexposureright{}}$ }
    \end{subfigure}
    \hfill
    \begin{subfigure}[b]{0.18\textwidth}
        \centering
        \includegraphics[width=\textwidth]{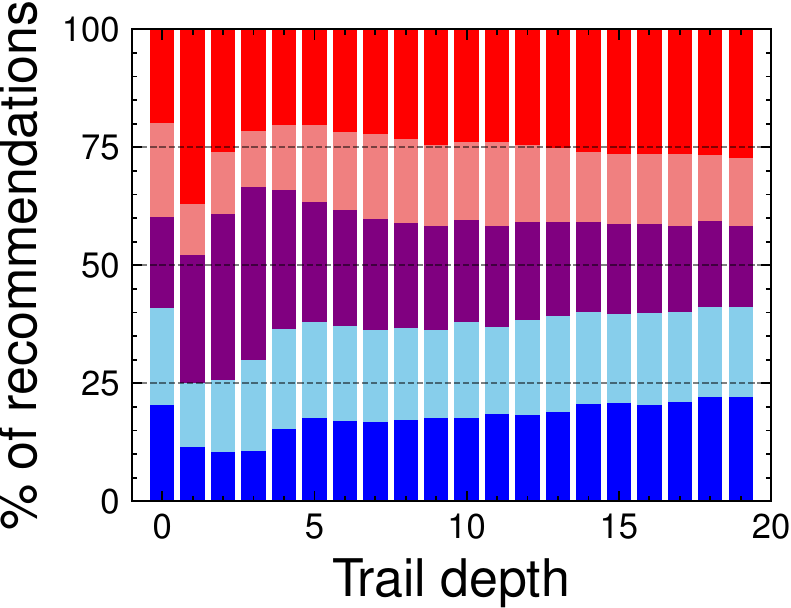}
        \caption{\rbin{} \\ $E^{\lbin{}}_{20} = \text{\leftexposurefarright{}}$\\ $E^{\rbin{}}_{20} = \text{\rightexposurefarright{}}$ }
    \end{subfigure}    
    \caption{Distribution of up-next recommendations for each sock puppet. The x-axis is the depth of the trail and the y-axis shows the distribution of up-next recommendations watched by each sock puppet at that depth. Depth 0 represents the seed video and is consistent across all the plots.}
    \label{fig:up-next-depth}
    \vspace{-.1in}
\end{figure*}

Figure \ref{fig:up-next-depth} plots the breakdown of ideology categories of up-next video recommendations for different sock puppet categories over trail depth up to 20 and the exposure value at that depth.
The colored stacked bar represents the distribution of the up-next recommendations, where each color corresponds to a particular ideology category.
The left-most column in each figure corresponds to depth 0: the uniformly randomly selected seed videos.
The uniform selection is reflected by the fact that the ideologies are equally distributed for each of the five sock puppet categories.
The immediately next column corresponds to a depth of 1 and so on.
For each sock puppet, we see a disproportionate ideological bias towards that sock puppet's ideology at different depths of the trail despite having the same starting seed video.
At depth 20, the \lbin{} and the \clbin{} sock puppets are exposed to \leftexposurefarleft{} and \leftexposureleft{} more left content respectively.
Similarly, the \rbin{} and \crbin{} sock puppets are exposed to \rightexposurefarright{} and \rightexposureright{} more right content respectively.
We note that the \rbin{} sock puppet has a slightly higher right exposure value ($E^{\rbin{}}_{20} = \text{\rightexposurefarright{}}$) than the left exposure value of the \lbin{} sock puppet ($E^{\lbin{}}_{20} = \text{\leftexposurefarleft{}}$). 
This suggests that the \rbin{} users have a slightly higher likelihood of being recommended and potentially watching an increasing number of right-biased content than the \lbin{} sock puppets.
Finally, we observe that the \cbin{} sock puppet sees a greater number of left-leaning content than it does right-leaning content.

YouTube's follow-up recommendations are increasingly biased when the user follows through on the recommendations regardless of the seed video.
Thus, the prior ideological bias of the user is likely to influence their up-next recommendations which increases the chances that the user will continue watching ideologically biased content.

\subsubsection*{RQ 3: Does following the recommendation trail lead to increasingly radical videos?}
\begin{figure}
    \centering
    \includegraphics{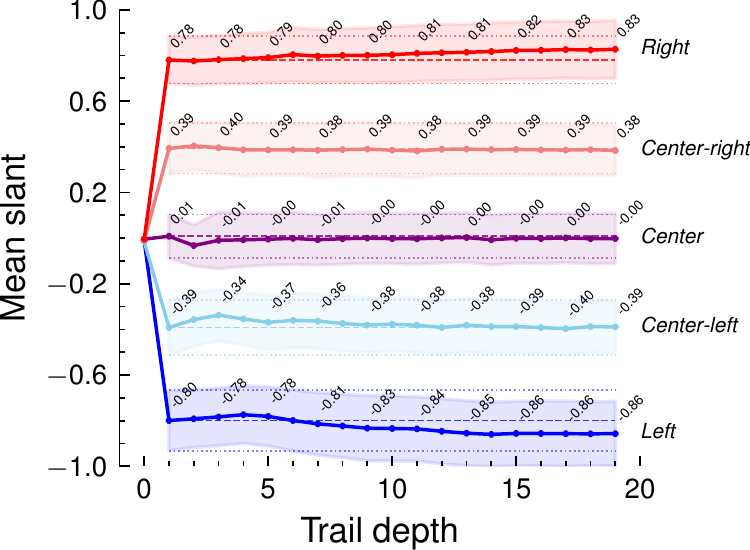}
    \caption{The mean slant of the ideologically same videos watched by each sock puppet ideology in the up-next recommendations. For example, the top red line corresponds to the mean slant of the \rbin{} videos watched by the \rbin{} sock puppet. The dotted and dashed lines correspond to 1-standard deviation and mean slant at depth 1 of the trail respectively. We can see the mean and standard deviation become progressively extreme for the \lbin{} and \rbin{} sock puppets.}
    \label{fig:radicalization}
\end{figure}

As we discussed in $\S$\ref{sec:estimating-video-ideology}, the estimated ideological slant of a video is within the range of -1 and +1.
Slants closer to -1 or +1 correspond to videos that are more ideologically extreme, i.e., radicalizing.
To determine whether videos deeper in the trail are more ideologically extreme, we take a more granular look at Figure \ref{fig:up-next-depth} by observing the continuous slant distributions within each ideology category.
To simplify, we only consider the slant distributions of ideology categories corresponding to the sock puppets' ideology.
Specifically, for the \lbin{} sock puppets we look at how the slant distribution for the videos within the \lbin{} ideology category vary as we traverse deeper into the recommendation trail.
We do this for all five ideology categories and show the mean and standard deviations of the distributions in Figure \ref{fig:radicalization}.

We see in Figure \ref{fig:radicalization} that the mean slant of the videos for the \lbin{} and the \rbin{} sock puppets and ideology gradually becomes more extreme as we traverse the trail.
For the \lbin{} ideology, it rises from -0.80 to -0.86 and for the \rbin{} ideology it rises from +0.78 to +0.83.
Due to how the slant is estimated, this suggests that the videos deeper in the recommendation trail were shared more by left or right audiences and are, therefore, more radical.
Looking at the standard deviations, we see that they also progressively move towards the extreme for the \lbin{} and the \rbin{} ideologies.

We do not see this progression for the \clbin{}, \cbin{}, and \crbin{} ideologies whose mean stagnates around the center for that ideology category.
This shows that radicalization occurs for only the \lbin{} and \rbin{} sock puppets, i.e., users who are already engaging with clearly slanted content. This key finding supports extant fears that it is precisely those YouTube users who are likely to progress to more ideologically radical and extreme videos in their YouTube sessions.

\section{Debiasing YouTube Recommendations}
\label{sec:intervention}
As our audit demonstrates, there is a pressing need to mitigate ideologically biased YouTube recommendations.
Unfortunately, YouTube is not self-incentivized to fix its recommendation system, which is designed to primarily optimize user engagement \cite{chen2019top}.
Thus, unlike the majority of prior research on \textit{top-down} debiasing interventions that typically aim to change the algorithm (e.g., by changing model architecture or the optimization function \cite{mehrabi2021survey}), we aim to design a \textit{bottom-up} intervention that gives users control over their recommendations without relying on cooperation from social media platforms \cite{kulynych2020pots}. 

In this section, we propose and systematically test a \textit{bottom-up} intervention framework by designing our debiasing approach, named \ToolX{}, that can be directly deployed by users (e.g., as a browser plug-in). 
There are two main challenges in designing such an intervention in a principled manner. 
First, since YouTube's recommendation algorithm is proprietary\footnote{Google/YouTube have sporadically published high-level descriptions of their recommendation algorithms \cite{zhou2010impact, covington2016deep, chen2019top} but not in sufficient enough detail to be of any practical use.} we do not possess exact knowledge of how the recommendation algorithm functions.
Second, as a direct consequence, our intervention can only rely on \textit{blackbox access} to YouTube's recommendation algorithm by querying it with an input (e.g., watching some videos) and observing the output (e.g., homepage recommendations). 
Given these challenges, we next describe the \ToolX{} framework to mitigate biased YouTube recommendations in a principled manner. 

\subsection{Proposed Approach}

\subsubsection{Overview}
 We design a Reinforcement Learning (RL) \cite{sutton2018reinforcement} based intervention approach to mitigate biased recommendations on YouTube. 
The intervention proceeds as follows. We train an RL model to minimize the recommendation of ideologically biased YouTube content by promoting more ideologically diverse, dissimilar, and balanced content on the user's homepage. As mentioned before, such diversity in information input is advantageous for social cohesion and citizen's attitudes, understanding, and knowledge \cite{helberger2018exposure}. 
The trained model \textit{obfuscates} the user's watch history by injecting suitable intervention videos if bias is detected on the user's homepage. 
We outline our RL methodology and key implementation details below. 

\subsubsection{Injecting intervention videos} 
Our intervention injects a fixed set of $m$ videos in \textit{batch mode}. 
More specifically, the intervention is employed at a time when the user is not actively using YouTube and so the set of intervention videos is introduced when the user is not watching any videos themselves.

We propose to train a model that operates on the user's YouTube homepage videos at any point in time and then chooses \textit{intervention} videos to inject to debias the homepage if the homepage is ideologically biased (i.e., if its mean ideological slant lies in the \rbin{} or \lbin{} ideology bins). To define this analytically, let $U$ denote the set of videos watched by the user before the intervention. Then, we can denote YouTube's recommendation algorithm as $\mathcal{A}$, which takes this watch history $U$ as input, and outputs a homepage $H$ consisting of a fixed number of $k$ videos.
Put simply, $H = \mathcal{A}(U)$.

At a high-level, the goal of the intervention is to \textit{modify} $U$ (the input to $\mathcal{A}$) by injecting $m$ intervention videos, such that the obtained homepage $H$ consists of more ideologically neutral, diverse, and/or dissimilar content so that it is balanced (i.e., mean ideology is close to 0).

\subsubsection{The goal of the RL model}
In the training phase, the RL model learns which videos to inject through \textit{trial and error}. By training over time, it learns to minimize errors in decision-making, by maximizing a function known as \textit{reward}. For our setting, we define this \textit{reward} function to measure how ideologically balanced homepage recommendations are (i.e., how close the average ideology slant of these recommendations is close to 0).

To do this, the reward function takes as input a user's homepage, and outputs a score between $-1$ and $0$, where $0$ denotes a perfectly balanced/diverse homepage, and $-1$ denotes a \textit{fully left} or \textit{fully right} homepage. The goal of our intervention is to obtain a homepage where the recommended videos achieve a reward value of close to 0, indicating that there is a balance of \lbin{} and \rbin{} videos and/or that the majority of recommended videos are ideologically moderate. We can now define such a reward function analytically using the slant scores of videos (Equation \ref{equation: slant}).

Let the homepage be denoted as $H$. We define the reward function as $\mathcal{O}(H) = -|\overline{S}(H)|$ where $\overline{S}(H)$ denotes the mean slant scores of the set of homepage videos $H$.

Again, the intervention injects a set of (\textit{intervention} videos) so as to maximize $\mathcal{O}(H)$, whose maximum value of 0 represents a homepage comprising of videos with \textit{mean} slant at center. If there are a large number of only left- or right-leaning videos, the reward value will be closer to $-1$.
Mathematically, our intervention introduces a set of intervention videos $I$ such that $\mathcal{O}(H')$ $\rightarrow$ 0, where  $H' = \mathcal{A}(U \, \Hat{\cup} \, I)$.\footnote{Note that the operator $\Hat{\cup}$ here denotes an ordered union between the set of videos, as the order in which videos are watched is important.}

While the reward function can be defined in different ways, our choice of $\mathcal{O}$ aims to minimize the mean ideology of a user's homepage. We select this reward function because it promotes ideological diversity. However, note that our approach is flexible enough to accommodate a variety of different such goals.

\subsubsection{Training the RL model}\label{training_rl}

Because we only have \textit{blackbox access} to YouTube's recommendation algorithm $\mathcal{A}$, we can only see the output homepage recommendations \textit{after} we inject a new video. Therefore, we design the learning framework as an \textit{iterative} process where each time step represents the addition of a new intervention video to eventually make up the complete intervention video set $I$. Such a framework is also known as a Markov Decision Process (MDP) \cite{bellman1957markovian}.

An MDP consists of the following components: \textit{states}, \textit{actions}, and the \textit{reward function}, which we detailed above. For our setting, each \textit{state} at a time step corresponds to the user's YouTube homepage at that time. Moreover, in an MDP, we can transition from one state to another using \textit{actions}, and the goal is to choose actions such that we move to states where the reward is maximized. In our case, we choose from 3 actions: to either watch a 1) \lbin{}-leaning video, 2) a \cbin{}-leaning video, or a 3) \rbin{}-leaning video. 

In short, the training process for our setting is summarized as:
\begin{itemize}
    \item For a learning \textit{episode}, the model starts with the user's pre-intervention homepage as the first state.
    \item Depending on this state, the model picks a particular video (\lbin{}, \cbin{}, \rbin{}) as the action to take.\footnote{The main motivation behind defining this action set is that each action should be able to send an explicit signal to the algorithm, to influence the recommendations. Thus, we only consider watching videos from the \lbin{}, \rbin{}, or \cbin{} ideology bins, as opposed to \clbin{} and \crbin{} videos.}
    \item The RL model goes back to the homepage and observes the recommendations to calculate the reward value using $\mathcal{O}$.
    \item Based on this reward obtained and the previous state, the model makes a decision for selecting the next action (i.e., the intervention video), and the process repeats from the first step.
    \item The learning \textit{episode} concludes when $m$ such videos have been introduced. We can hence train the RL model for many such episodes till it has learned to achieve desired reward, and can debias YouTube homepages effectively.
    \item In this manner, this trained RL model has learnt an \textit{optimal policy} that allows it to \textit{map} each state (homepage) to the best possible action (video to inject) for that time step.
\end{itemize}

\begin{wrapfigure}{R}{0.6\textwidth}
    \centering
    \includegraphics[width=0.6\textwidth]{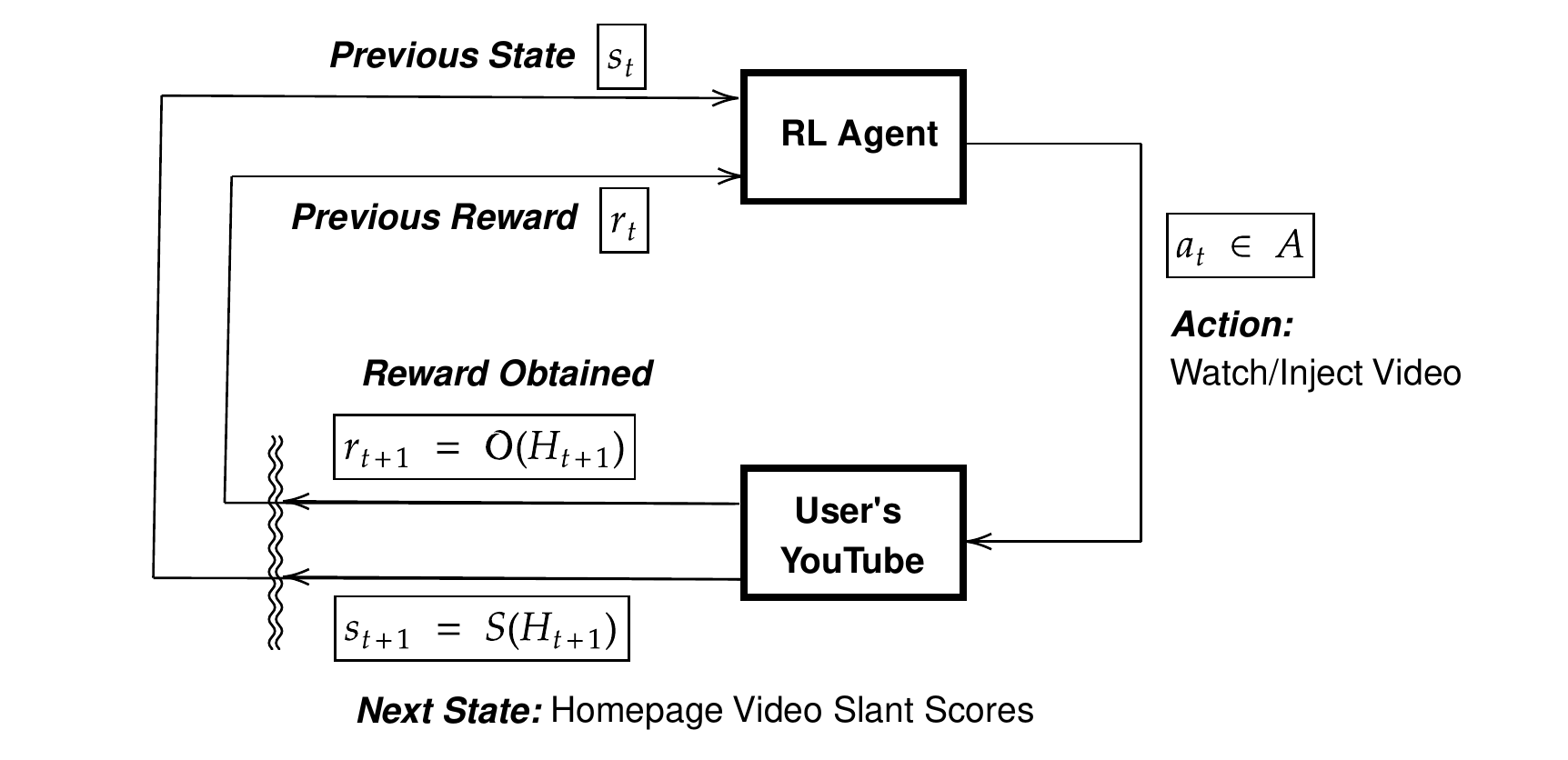}
    \caption{The proposed RL based approach.}
    \label{fig:rl}
\end{wrapfigure}

We will now describe some of these terms and formulations analytically.

\vspace{.05in} \noindent \textbf{States.}
At each time step $t$, the state $s_t$ is defined as a $k$ length vector of slant scores computed on the set of homepage $H_t$ videos. We denote this as $s_t = S(H_t)$. Thus, our states encode the political ideology of the homepage at that time step.

\vspace{.05in} \noindent \textbf{Actions.}
At each time step $t$, the model interacts with the blackbox environment (the YouTube recommendation algorithm $\mathcal{A}$) by selecting one of the actions from the set of allowed actions $A = \{a_L, a_C, a_R\}$ where $a_L, a_C, a_R$ refer to watching a \lbin{}, \cbin{}, or \rbin{} video, respectively. As mentioned before, we opt for this action set since each of these actions will send an explicit signal to the recommendation algorithm $\mathcal{A}$.

\vspace{.05in} \noindent \textbf{Reward.} At time step $t$, upon selecting a possible video injection $a_t$ from $A$, the agent receives the set of next state observations $s_{t+1}$ from the YouTube environment, as well as a scalar reward $r_{t+1}$. As described before, $s_{t+1}$ is denoted as the slant scores of the homepage videos $H_{t+1}$. Since we want to learn a policy that maximizes the reward value, the immediate reward $r_{t+1}$ is simply defined as $r_{t+1} = \mathcal{O}(H_{t+1}) = -|\overline{S}(H_{t+1})|$.

\vspace{.05in} \noindent \textbf{Model architecture.}
Using this formulated MDP, we learn an optimal policy $\pi$ that maps each state to an action using RL algorithms. 
Specifically, we utilize the well-known Deep Q-Network \cite{mnih2013playing} (DQN) RL algorithm. There are many benefits to using DQN and its extensions. First, the DQN algorithm is \textit{simplistic}, and complements our MDP representing a simple/limited state-action space. Second, the approach is \textit{model-free} and \textit{off-policy}, which does not require us to explicitly know the blackbox YouTube environment, and instead allows us to learn $\pi$ using samples obtained from exploring the environment itself. 
Finally, combined with extensions such as Prioritized Experience Replay (PER) \cite{schaul2015prioritized} and Dueling-DQN \cite{wang2016dueling}, the DQN architecture is stable and converges quickly. We employ these extensions in our experiments as well.\footnote{Because our state space consists of $k$ length slant vectors, we use a Multi-layer Perceptron (MLP) Neural Network (2 layers with 64 neurons each) as our function approximator, as opposed to a more complex Convolutional Neural Network (CNN) used in the original paper \cite{mnih2013playing}.}
This description of the MDP and the RL based intervention framework is also illustrated in Figure \ref{fig:rl}.

\subsubsection{Additional details}

\noindent \textbf{Left-leaning and right-leaning models.}
We implement the intervention framework by training RL models for two types of users: left-leaning and right-leaning comprising of the \lbin{} and \rbin{} ideology bins respectively. We consider these two bins as they exhibit the most ideological bias in recommendations (as evidenced via the audit). Thus, given the homepage prior to the intervention, if the mean slant of the homepage (videos) lies in the \rbin{} ideology bin we invoke the RL model for right-leaning users otherwise if it lies in the \lbin{} ideology bin we use the RL model for left-leaning users. 

There are many benefits to training \textit{specialized} models as opposed to training a \textit{general} model that can select intervention videos for users of any ideological leaning. A general model would have to first \textit{learn} to distinguish between left-leaning and right-leaning users, whereas this information is encoded explicitly prior to training the specialized models. Note that this is trivial to do, since we can check whether $\overline{S}(H)$ is positive or negative prior to invoking the intervention. Other benefits to having specialized models include:
(1) we can analyze each model's decisions more clearly since we know what they should intrinsically optimize for, 
(2) the MDP is less complex as each model is exposed to fewer states, and 
(3) models converge more quickly as the learning problem is simplified. This may be the biggest advantage -- RL models are sample inefficient \cite{yarats2021improving} and often require hundreds of thousands to millions of learning episodes to train well. As we cannot ``simulate'' YouTube offline, we have to resort to first training sock puppets to mimic real users, then watch videos and observe the homepage, which makes training an RL agent for such a large number of episodes not tractable.  Thus, simplifying the MDP (state-action space) for specific types of users allows us to effectively train an RL model in fewer episodes. 

\vspace{.05in} \noindent \textbf{Experiments for training and testing.} We train the RL models by utilizing sock puppets as in $\S$\ref{sec:audit}. We begin a training episode by first training a sock puppet of the ideology corresponding to the particular RL model (i.e., left-leaning or right-leaning). 
As part of this process, each sock puppet ``watches" 100 videos of the corresponding ideology; that is $|U| = 100$. 
Further, to analyze how many intervention videos are needed to debias the homepage efficaciously, we vary the parameter $m$ from $25, 50, 75,$ to $100$ videos\footnote{Note that we do not need to train separate RL models for different values of $m$ as we can train a model for any $m'$ and still utilize it for $m$ intervention videos. Upon deployment, we can keep injecting videos sequentially till $m$ videos are introduced.}.
The training episode concludes when $m$ videos have been injected, as described in $\S$\ref{training_rl}. In this manner, the RL agent explores this state-action-reward space by consistently learning over 2500 episodes, and learns an optimal policy $\pi$ that maps states to actions.  We find that the model starts achieving desirable reward values after training for this duration.

While we only train the models to debias homepage recommendations, we \textit{test} the intervention out on both the homepage as well as up-next recommendations. We carry out these analyses similar to $\S$\ref{sec:audit} by utilizing sock puppets and use the same metrics for measuring ideological bias. In total, we employ $1894$ sock puppets for these experiments which correspond to $1,589,105$ YouTube videos.


\subsection{Results}

\subsubsection{Debiasing homepage recommendations}\label{sec:homepage_recs}
We analyze whether the intervention framework debiases a user's homepages and how many intervention videos need to be injected. 
To this end, we collect homepage video recommendations for both \rbin{} and \lbin{} users (sock puppets), \textit{before} and \textit{after} the intervention and analyze the distributions of video slant scores seen by the sock puppets at both times to test whether they are \textit{statistically significantly} different. 
Further, we also vary the value of $m$, i.e., the number of intervention videos injected from $25, 50, 75$ to $100$ videos to ascertain how adding more intervention videos affects the extent to which the homepages is debiased.

\begin{figure*}[ht]
    \centering
    \begin{subfigure}[b]{0.45\textwidth}
        \centering
        \includegraphics[scale=0.65]{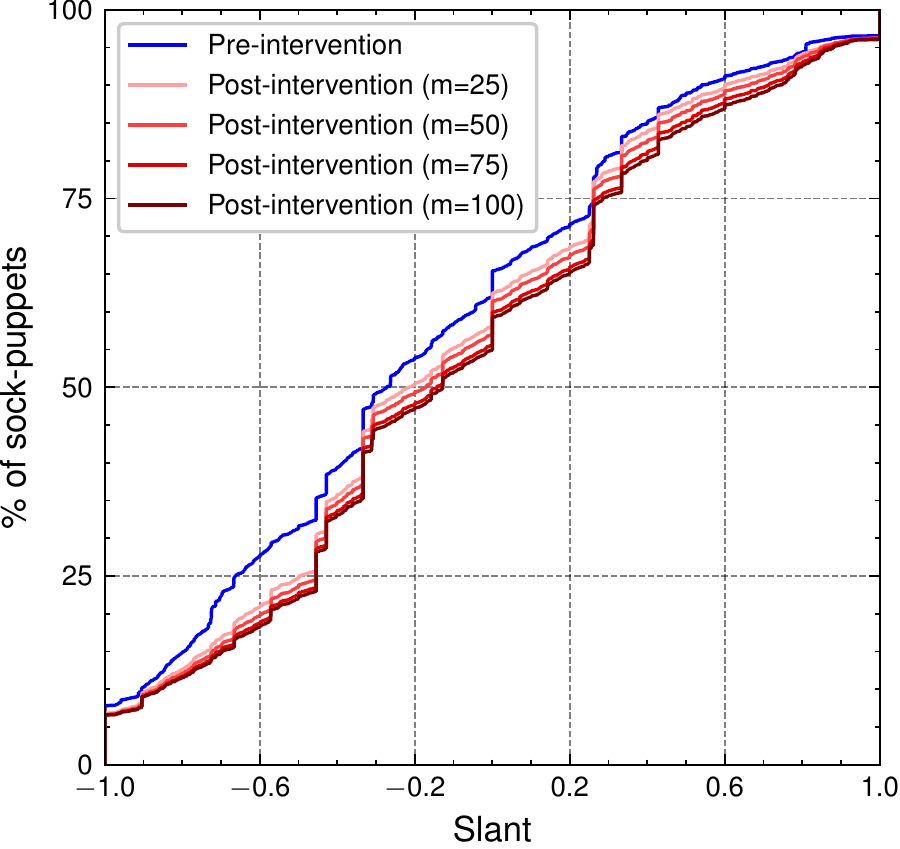}
        \caption{CDFs for left-leaning model.}
        \label{fig:left_rl_cdf}    
    \end{subfigure}
    \hfill
    \begin{subfigure}[b]{0.45\textwidth}
        \centering
        \includegraphics[scale=0.65]{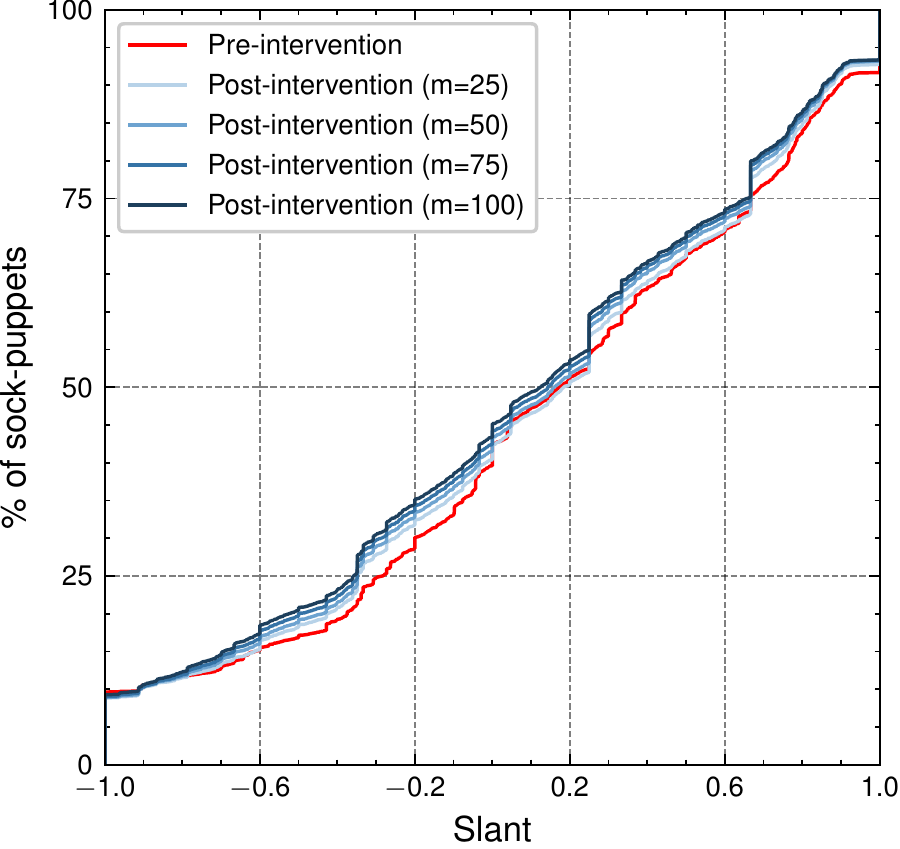}
        \caption{CDFs for right-leaning model.}
        \label{fig:right_rl_cdf}
    \end{subfigure}
    \caption{CDFs showcasing the slant distribution for sock puppets pre-intervention and post-intervention (for different values of $m$, i.e., the \# of intervention videos injected).}
    \label{fig:rl_cdf}
\end{figure*}

\begin{wraptable}{r}{0.59\textwidth}
\vspace{-.1in}
\centering
\caption{Statistic values for the Kolmogorov-Smirnov test conducted between the pre-intervention distribution corresponding to the model (right-leaning/left-leaning) and different post-intervention distributions obtained by varying $m$. The * indicates statistical significance as all $p$-values = 0.}
\label{tab:ks-cdf-rl}
\begin{tabular}{|ccc|}
\hline
Distribution & \begin{tabular}[c]{@{}c@{}}Left-leaning \\ model\end{tabular} & \begin{tabular}[c]{@{}c@{}}Right-leaning \\ model\end{tabular} \\ \hline
Post-intervention ($m=25$) & 0.072* & 0.036* \\
Post-intervention ($m=50$) & 0.084* & 0.046* \\
Post-intervention ($m=75$) & 0.093* & 0.055* \\
Post-intervention ($m=100$) & 0.098* & 0.063* \\ \hline
\end{tabular}%
\end{wraptable}

The results are presented as Cumulative Distribution Functions (CDFs) for both the left-leaning and right-leaning model in Figure \ref{fig:left_rl_cdf} and Figure \ref{fig:right_rl_cdf}, respectively. The visual difference between the pre-intervention and post-intervention curves for both models indicates that the intervention does lead to a change in the ideological distribution of homepage videos. This \textit{visual gap} between the pre-intervention and post-intervention curves is more pronounced for \lbin{} sock puppets than for \rbin{} sock puppets. Finally, increasing $m$ leads to more debiasing as the pre-intervention and post-intervention curves become visually more distinct upon doing so.

To quantify these observations, for the left-leaning and right-leaning model, we conduct the Kolmogorov-Smirnov (KS) test between the pre-intervention distribution and each of the post-intervention distributions we obtain by varying $m$. Table \ref{tab:ks-cdf-rl} shows that for both models the pre- and post intervention distributions are all statistically significantly different because the $p$-values $= 0$ and the test statistic is greater than 0. Moreover, for the same value of $m$, the test statistic for the left-leaning model is higher than that for the right-leaning model. This indicates that it is \textit{indeed easier to debias a \lbin{} user than a \rbin{} user}. Finally, Table \ref{tab:ks-cdf-rl} also shows that by increasing $m$, the test statistic values increase, indicating that introducing a higher number of intervention videos debiases the homepage to a larger extent\footnote{The medians of the distributions also illustrate this. For the left-leaning model the median goes from $-0.272$ (pre) $\rightarrow -0.171$ (post, $m=50$) $\rightarrow -0.128$ (post, $m=100$). For the right-leaning model the median goes from $0.173$ (pre) $\rightarrow -0.158$ (post, $m=50$) $\rightarrow -0.122$ (post, $m=100$).}. However, even for $m=25$ the intervention is effective as the KS test statistic obtained is statistically significant.

Next, we analyze the ideological leaning of recommended homepage videos before and after the intervention to assess if the intervention led to more ideologically diverse and balanced recommendations. Table \ref{tab:bin-dist-cdf} shows that for the left-leaning model, this is the case as the \% of \lbin{} videos decreases from $51.23\%$ to $37.34\%$ ($m=100$), and both the \rbin{} video \% and \cbin{} video \% increase significantly-- from $16.87\%$ to $26.67\%$ ($m=100$), and from $31.90\%$ to $35.99\%$ ($m=100$), respectively. For the right-leaning model, although there is a similar trend of increase in diverse content, the \rbin{} video \% does not decrease much ($44.73\% \rightarrow 42.40\%$), and the \cbin{} video \% also decreases only slightly ($31.66\% \rightarrow 28.41\%$) to account for the increase in \lbin{} video \% from $23.61\%$ to $29.19\%$ ($m=100$). Thus, overall, more ideologically diverse and balanced content is recommended post the intervention for both models, even though the changes are less pronounced for right-leaning users.

\begin{table}[ht]
\centering
\caption{Percentage of \lbin{}, \rbin{}, and \cbin{} homepage videos encountered pre-intervention and post-intervention for both the left-leaning and right-leaning models.}
\label{tab:bin-dist-cdf}
\begin{tabular}{|c|c|ccc|}
\hline
Model Type & Distribution & \lbin{} video \% & \rbin{} video \% & \cbin{} video \% \\ \hline
\multirow{5}{*}{\begin{tabular}[c]{@{}c@{}}Left-leaning\\ model\end{tabular}} & Pre-intervention & 51.23 & 16.87 & 31.90 \\
 & Post-intervention ($m=25$) & 42.72 & 21.23 & 36.05 \\
 & Post-intervention ($m=50$) & 40.62 & 22.99 & 36.39 \\
 & Post-intervention ($m=75$) & 38.59 & 25.20 & 36.21 \\
 & Post-intervention ($m=100$) & 37.34 & 26.67 & 35.99 \\ \hline
\multirow{5}{*}{\begin{tabular}[c]{@{}c@{}}Right-leaning\\ model\end{tabular}} & Pre-intervention & 23.61 & 44.73 & 31.66 \\
 & Post-intervention ($m=25$) & 25.96 & 45.91 & 28.13 \\
 & Post-intervention ($m=50$) & 26.99 & 44.51 & 28.50 \\
 & Post-intervention ($m=75$) & 28.16 & 43.35 & 28.49 \\
 & Post-intervention ($m=100$) & 29.19 & 42.40 & 28.41 \\ \hline
\end{tabular}
\end{table}

\subsubsection{Debiasing up-next recommendations}

\begin{wraptable}{r}{0.57\textwidth}
\vspace{-.1in}
\centering
\caption{Exposure metric for pre-intervention and post-intervention up-next recommendation trails.}
\label{tab:exposure-rl}
\begin{tabular}{|c|c|c|}
\hline
Model Type & Distribution & Exposure \\ \hline
\multirow{2}{*}{Left-leaning} & Pre-intervention & $E^{\lbin{}}_{20} =  0.7183$ \\
 & Post-intervention & $E^{\lbin{}}_{20} = 0.7140$ \\ \hline
\multirow{2}{*}{Right-leaning} & Pre-intervention & $E^{\rbin{}}_{20} = 0.6607$ \\
 & Post-intervention & $E^{\rbin{}}_{20} = 0.5138$ \\ \hline
\end{tabular}%
\end{wraptable}
As mentioned before, while we train the RL models to debias homepage recommendations, we also analyze the effect of the intervention on up-next recommendations. To do so, we utilize the exposure metric (Equation \ref{exposure_eq}). We use the same video seed as the ideological leaning of the sock puppet (\lbin{} or \rbin{}) and then compute exposure for recommendation trails of depth 20 (as in $\S$\ref{sec:audit}), pre- and post-intervention. As we found that increasing $m$ improves debiasing of homepage recommendations, we utilize $m=100$ for these experiments but note that other (smaller) values of $m$ can also be used.

Table \ref{tab:exposure-rl} shows the exposure results. For both models, exposure decreases after the intervention, indicating that the intervention can also effectively debias up-next recommendations. Surprisingly, for the left-leaning model the pre-intervention exposure value does not decrease much ($0.7183 \rightarrow 0.7140$) and is more pronounced for the right-leaning model: from $0.6607$ before to $0.5138$ after the intervention. Future work should study more principled intervention for the up-next recommendation trails.

\subsubsection{Interpreting the RL model's decisions}
It is important that the RL model's decisions are transparent to the user of our intervention. 
The RL model is not readily interpretable because it utilizes deep neural networks for function approximation, which does not lend itself to providing meaningful insight to users about how the intervention is changing their homepage recommendations. 
%
To explain the RL model's decision, we utilize Decision Trees (DTs) for post-hoc interpretability. More specifically, we use Decision Trees (DTs) to explain the RL model's decisions primarily due to their ease of visual interpretability. They are essentially a visual descriptor of a rule set and are commonly used in machine learning research \cite{gunning2019darpa}.

\begin{figure}[ht]
    \centering
    \includegraphics[width=0.55\textwidth]{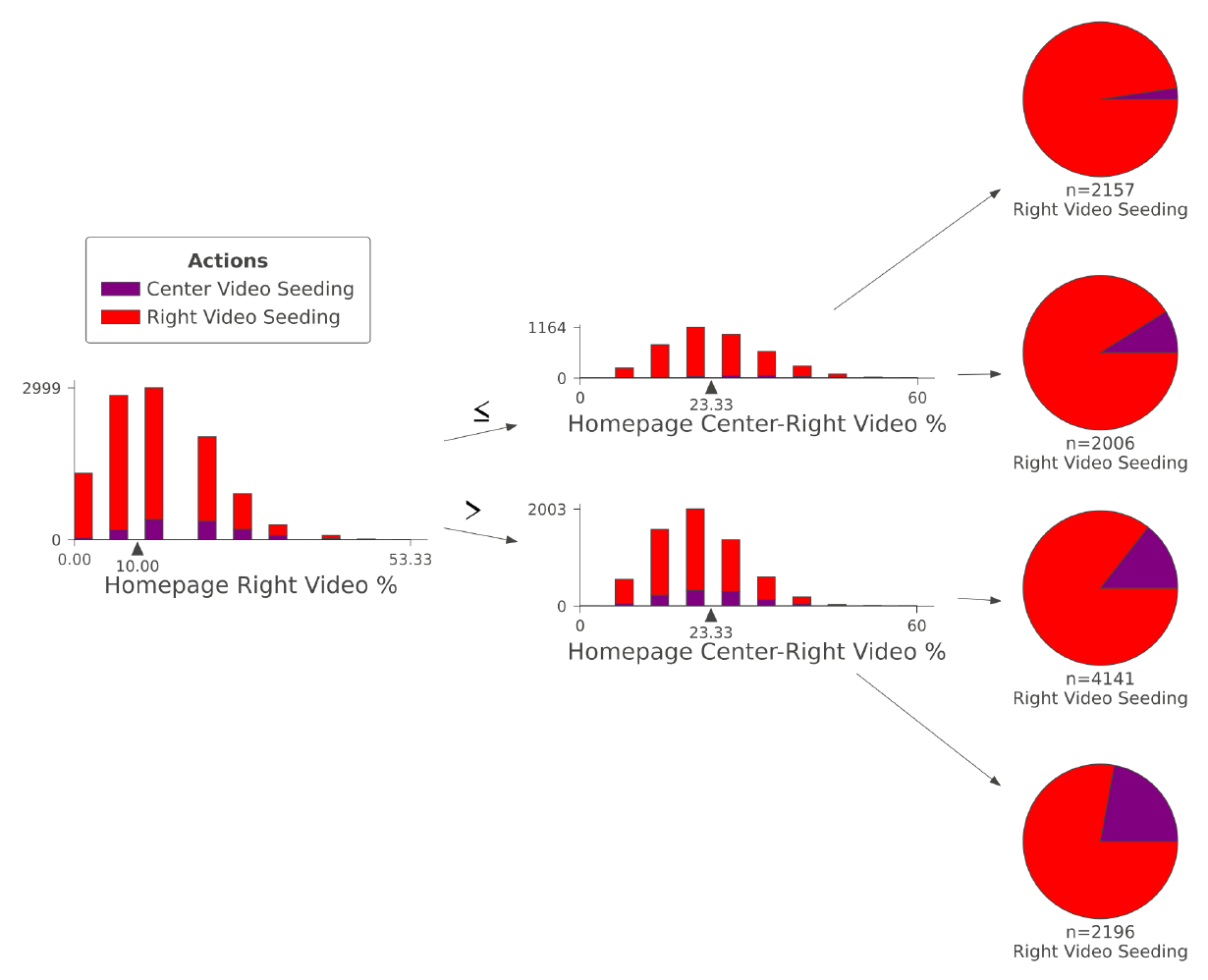}
    \caption{DT proxy model of the left-leaning RL model: (1) Read from left-to-right; the upwards branch of the tree from a node corresponds to the condition at the node being false, and the downwards branch of the tree corresponds to the condition being true. (2) Intermediate nodes are represented via histograms and the leaf nodes via pie charts. The histogram maps the state-action space-- thus corresponding to the feature of the dataset and the threshold value that leads to the largest information gain (least entropy). The pie charts show the distribution of the actions taken by the model when the corresponding sequences of rules in previous nodes are met such that we reach that particular leaf node. (3) The colors represent the action chosen: purple for \cbin{} injection ($a_C$) and red for \rbin{} injection ($a_R$). It can be seen that the model considers the \% of right-leaning videos on the homepage, and then picks \rbin{} injection a majority of the time. Intuitively, \cbin{} injection is picked with higher frequency if the model sees that there are a large number of right-leaning videos on the homepage.}
    \label{fig:dt_l}
\end{figure}
\begin{figure}[ht]
    \centering
    \includegraphics[width=0.55\textwidth]{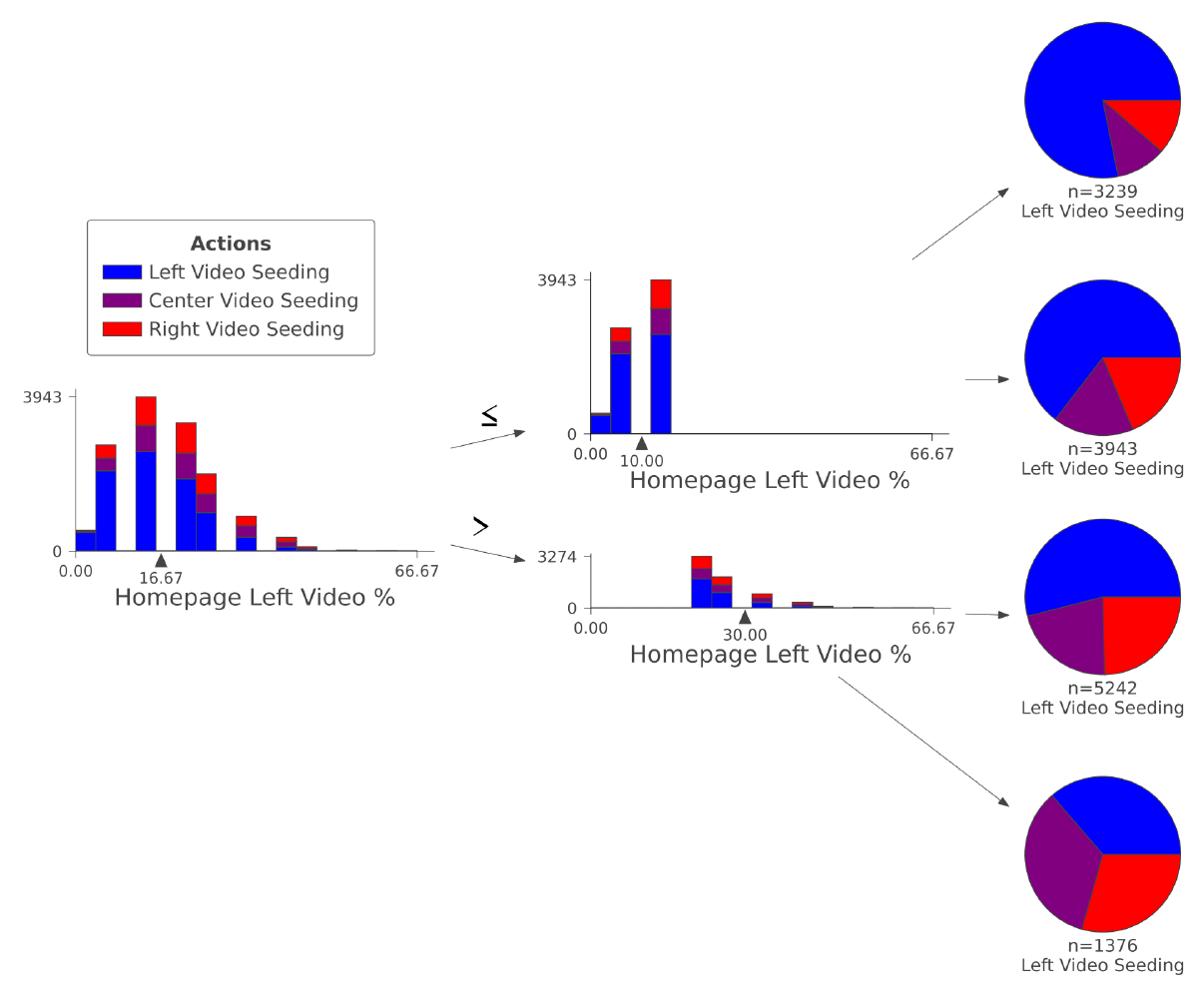}
    \caption{DT proxy model of the right-leaning RL model: To be read similarly to Figure \ref{fig:dt_l}. Note that here the model considers the \% of left-leaning videos on the homepage, and then picks \lbin{} video injection a majority of time. Infrequently \cbin{} and \rbin{} seeding are also picked to possibly offset extreme recommendations on the homepage.}
    \label{fig:dt_r}
\end{figure}

We utilize the state-action pairs obtained as a result of our previous experiments with the intervention framework and train a DT classifier on these data\footnote{Thus, this is a classification problem where states constitute the dataset and the actions constitute the class to predict.}. The results for the left-leaning and right-leaning models are shown in Figure \ref{fig:dt_l} and \ref{fig:dt_r}, respectively. The figures show that left-leaning model picks \rbin{} injection a majority of the time and the right-leaning model picks \lbin{} injection most frequently. Other classes/actions for both models are generally picked less frequently. 

However, at times the choice of picking the other actions is also intuitive. For instance, for the left-leaning model if the \% of \crbin{} videos on the homepage is greater than a certain threshold the model chooses \cbin{} seeding more frequently than it did before. Similarly, for the right-leaning model if the \rbin{} video \% on the homepage goes above a threshold value, the model tends to pick the other actions (such as \cbin{} or \rbin{}\footnote{This is reasonable as the \cbin{} ideology bin has a slight left bias (as mentioned in $\S$\ref{sec:audit}, the \cbin{} slant distribution median is at $-0.04$), and hence, it makes sense that the model might pick \rbin{} injection (albeit, infrequently) when there are a large number of \lbin{} videos being recommended on the homepage.} injection) to offset this overtly left skew in recommendations. 

\section{Discussion}
\label{sec:discussion}

Algorithmic recommendation systems of social media platforms are solely designed to optimize user engagement. In the case of political content, there are serious concerns that this incentive leads to the promotion of, exposure to, and engagement with content that is politically like-minded, reinforcing users' biases, and that is progressively more radical, putting users in the oft-described ``rabbit hole'' of radicalization.

Our work aimed to offer a holistic understanding of whether and how algorithmic recommendation systems of YouTube -- the most popular social media platform -- drive exposure to ideologically congenial and progressively more radical content and -- more importantly -- how to design interventions that can effectively mitigate the potential ideological bias in these systems.

We offer three key findings with clear theoretical, methodological, and practical implications:

First, we find clear ideological bias in recommendations stemming from the user's prior exposure. This bias exists in both homepage and up-next recommendations, where the recommended videos are disproportionately likely to align with the ideology of the user (addressing RQ1). As the user traverses the up-next recommendation trail, the number of biased recommendations also increases exposing the user to increasingly biased content (RQ2). Furthermore and most troubling is is the finding that the number of biased videos at higher depths is not only greater but also that the recommended videos are also increasingly radical (RQ3). In other words, following the trail of up-next recommendations (akin to a user's decision to watch the videos that YouTube recommends, and -- after all -- 70\% of YouTube's content is watched from recommendations \cite{rodriguez2018ytrecommendationsQZ} without requiring any additional effort, action, or decision from the users) leads to a higher number of biased content that is also more radical. While the user can choose to watch an unbiased recommendation, they are more likely to select a video recommendation that aligns with their prior beliefs, especially if the user is already partisan, as suggested by the selective exposure framework  \cite{stroud2010polarization}. This reinvigorates the loop effect and continues the vicious cycle of biased and radicalizing content. Our sock puppet based audit, examining three consequential outcomes -- ideological bias, its magnitude, and radicalization -- in both homepages and up-next recommendations reconciles conflicting evidence in prior literature on the role of YouTube's recommendation algorithm in perpetuating bias. Through the use of trained sock puppets, we rule out external factors to radicalization and demonstrate how prior user watch history is an important step to algorithmic radicalization.

Second, it is possible and feasible -- without relying on cooperation from YouTube -- to debias its recommendation algorithm toward more balanced and diverse recommendations. Our RL-based intervention framework is able to reduce the bias in a user's homepage recommendations by obfuscating their watch history when the user is not active on YouTube. Our intervention approach can effectively debias \lbin{} and \rbin{} user's homepages as well as up-next recommendations by increasing exposure to more diverse and balanced content. Furthermore, we show how to interpret the actions of the model -- i.e., why it made a particular decision about which video to watch -- to make it more transparent to end users. 

Our last key finding regards the presence of right-wing bias in YouTube's recommendation algorithm. Our audit shows that the \rbin{} users are more likely to encounter more right-leaning recommendations than any other ideological group examined in our work. Although this finding is only suggestive, as pronounced differences between right-leaning and other users did not emerge across all of our models and outcomes variables, this finding is nevertheless important in the current polarized political climate in the U.S., and especially as some social media platforms (e.g., Twitter) were found to be promoting right-leaning content \cite{huszar2022algorithmic} and Republicans are twice as likely as Democrats to see right-leaning political videos on YouTube \cite{pew2020youtubenews}. 
Furthermore, and in line with this finding, we also show that it is substantially more challenging to effectively debias the YouTube homepages of right-leaning users than the homepages of left-leaning YouTube users using our framework.

Next, we address some limitations of our work and also some ethical and legal considerations related to \ToolX{}.

\subsection{Limitations}
Despite the robust and systematic evidence, our design and evaluation are not free from limitations.  

First, it is not possible for our audience-based method to estimate slant of videos that have not been shared on Twitter. This limitation applies to all such methods that rely on analyzing Twitter sharing patterns.\footnote{For the \numuniqtrainingvideos{} unique videos used during training, we were able to estimate slants for \numtrainingslants{} of those videos yielding \trainingcoverage{} coverage. During testing, our \numsockpuppets{} sock puppets encounter \numuniqtestingvideos{} unique videos for which we are able to estimate the slant of \numtestingslants{} videos yielding \testingcoverage{} coverage. For the remaining videos, we could not find any tweets for \uncoveredslantnotweets{}, found tweets but could not identify landmarks for \uncoveredslantnolandmarks{}, and found tweets but could not identify sufficient landmarks for \uncoveredslantnotenoughlandmarks{}.} This limitation, although concerning, does not demerit the automated slant estimation method and its usefulness especially when compared to prior work that used a small number of manually-assigned channel labels \cite{ribeiro2020auditing} which, as discussed in $\S$\ref{sec:estimating-video-ideology}, can be misleading. Future work can incorporate automated or crowd-sourced content-based methods to estimate slant of the remaining videos. 
%

Second, we note that recommendation algorithms are constantly changing. We implicitly assumed that YouTube's recommendation algorithm $\mathcal{A}$ is \textit{static} and does not learn in an online fashion simultaneously as we select the intervention video set $I$. Without this assumption, the learning problem becomes significantly more complex, as it becomes a \textit{multi-agent} RL problem. Unlike an MDP formulation where a state-action pair would map to the same state, in this setting, state transitions can be non-stationary and change with time. Due to this, we can have exponentially increasing state-action spaces, further complicating the problem. Algorithms for solving multi-agent RL problems are also not as efficacious as single-agent RL algorithms, and might not quickly converge to feasible solutions.

Lastly, we acknowledge that users who are most in need of our intervention (i.e., those already extreme, those at high risk of radicalization, and so forth) are also the least likely to wish to debias their recommendations and install any plug-ins offered by the scientific community. In other words, those most susceptible to problematic and biased recommendations might not employ such mitigation tools. We note that our intervention does not require users to watch the intervention videos. After all, a user will not voluntarily watch several intervention videos that challenge their views \cite{stroud2010polarization}. Our tool, therefore, is designed to perform the required video watches in a background tab during a time when the user is not directly interacting with YouTube. Nevertheless, this cannot happen without an explicit informed consent from the user and the user's willingness to install our plug-in. Further efforts are needed to examine how such bottom-up interventions can be popularized and adapted. We offer foundational evidence and applicable tools, and leave this key challenge for future research.


\subsection{Ethical Considerations}
We now discuss a few ethical and legal considerations related to our intervention by outlining its potential harms and benefits for YouTube, users, and society at large. Because \ToolX{} is designed to be directly employed by users without needing any explicit cooperation with YouTube, it represents a computer-supported collective action effort \cite{shaw2014computer}. 

%

When it comes to YouTube, its recommendation system is optimized to maximize user engagement which directly translates into advertising revenue \cite{youtubehowmoney}. Therefore, tweaking YouTube's recommendations might result in less user engagement, inadvertently decreasing the advertising revenue of the content publishers whose videos will be less likely to get recommended. However, to the extent that the increased revenue comes from time spent watching extreme or radicalizing videos, our intervention may more significantly impact the creators of this kind of problematic content, which is not necessarily harmful to the overall content creator ecosystem and may minimize the negative societal effects of otherwise preventable exposures.


When it comes to YouTube users, our intervention might result in less relevant recommendations. However, we believe it is justifiable since our intervention is expected to be used by consenting users and also because ``giving people more of what they want'' is not necessarily inherently beneficial to the users or society at large. 

On that note, our intervention is designed to expose users to ideologically balanced, diverse, and/or dissimilar YouTube videos. Such exposure, as aforementioned, is crucial to bridge the ever increasing political divides in the U.S. \cite{luther2021partisanship}, in that partisans need to encounter diverse perspectives on key political issues and such encounters are impossible within the confines of recommendations to increasingly biased and radical content.

\subsection{Legal Considerations}

YouTube's Terms of Service (TOS) \cite{youtubeTOS} allow the use of the service as long as the user agrees to certain terms. Any violation of the TOS of a website could be considered a violation of the Computer Fraud and Abuse Act (CFAA, 18 U.S. Code § 1030). In other words, violations of TOS such as unauthorized access or exceeding authorized access to a protected system might be penalizable under CFAA. While YouTube is considered a protected system under CFAA, it remains unclear whether our auditing and intervention approaches engage in unauthorized access  \cite{mackey2021vanburen}. As note below, overly broad interpretations that attempt to equate TOS violations to crimes under CFAA have been turned down by federal courts in the past.

In terms of the legality of audit by researchers, YouTube's TOS restrict access to the service via ``automated means'', which means that our sock puppet based auditing system is in violation of YouTube's TOS. Such violations are subject to penalty under the CFAA because they involve probing a computer system in a misleading manner \cite{fiesler2020no}. 
To remove overbroad interpretations of the CFAA, the case of Sandvig v. Barr \cite{aclu2019CFAA} attempted to restrict the use of CFAA to criminally prosecute researchers who violate TOS based on the argument that it violates their ``First Amendment right to engage in harmless false speech''.
In 2020, a federal court ruled that the violation of a website's TOS is \textit{not }grounds for being held criminally liable under the CFAA \cite{naomi2020fedtos}.
This landmark decision provides legitimacy to sock puppet based auditing of algorithmic systems, which is now considered a well-established methodology in the academic research community \cite{ribeiro2020auditing,papadamou2020just,papadamou2020over} 

In terms of legality of the intervention by users, YouTube does not allow users to fraudulently engage with its service in any manner. As our intervention introduces external, non-recommended videos in a user's watch history, it can be construed as fraudulent behavior on part of the user in how they interact with the service. However, YouTube is a public system and users are ``authorized'' to watch any publicly available video on the platform \cite{youtubemission}.
There is legal precedent that authorized access to YouTube, even in violation of TOS, does not violate the CFAA. In Van Buren v. United States \cite{mackey2021vanburen}, the court ruled that a police officer who accesses a law enforcement database in exchange for money did not violate the CFAA. 
The rationale by the court was that because the officer was already authorized to access the system, their actions do not violate CFAA.
Based on this ruling, we argue that the users of our intervention cannot be held liable under the CFAA because they do not exceed in their authorization to access publicly available YouTube.

\section{Concluding Remarks}
Rabbit holes of radicalization in algorithmic recommendation systems are detrimental to democracy, leading to extremity and emboldening some users to engage in unlawful actions, as was the case of January 6th riot on the Capitol. 
In contrast, exposure to diverse and familiarity with various perspectives is a \textit{sine qua non} of an effectively functioning society. 
Considering the current political climate in the US and internationally, where citizens are more divided than ever and where support for democratic norms is on decline \cite{kalmoe2019lethal}, auditing the recommendation algorithms of social media platforms for ideological bias and proposing principled solutions to minimizing it is both timely and needed.

Our audit joins prior work \cite{hussein2020misinformation,sanna2020yttrex,papadamou2020just,papadamou2020over} that uncovers ideological bias in YouTube's recommendation systems. 
Our large-scale experiments find that recommendations that users receive \textit{are} aligned with their ideology, and this is especially true for right-leaning users, who \textit{are} recommended an increasing number of ideologically congenial videos, also especially among the political right, and moreover that the recommendations \textit{are} progressively more extreme, leading to the claimed rabbit holes \cite{tufekci2018youtube}. 

On a more optimistic note, we show that it is possible to mitigate thees ideological biases. Our proposed intervention indeed leads YouTube's algorithm to recommend more ideologically neutral, diverse, and dissimilar video recommendations, leading to a more ideologically balanced exposure. The next step in this work is to develop an open-source intervention tool \ToolX{}, which -- building on our intervention -- is trained to automatically estimate bias in YouTube recommendations and seamlessly, in the background, inject videos to mitigate the bias in a principled manner. The groundwork laid out here demonstrates that this approach can ideologically debias homepage and up-next recommendations, and that the tool's actions are transparent and can be interpreted by end users. Importantly, the \ToolX{}'s framework can be readily adapted to help users control and mitigate a wide range of undesired recommendations (e.g., misinformation, click-bait, hate speech, etc.) and its infrastructure can be adapted to other social media platforms. 

We note that a widespread adoption of \ToolX{} among the most pressing target (i.e., the extreme and strongly partisan citizens) is unlikely. After all, those citizens may not be aware of their own biases and the biases in their social media feeds, and -- even if aware -- unwilling to change them. 
That said, recent work \cite{huang21poisoning} has shown that even a limited number of users can ``poison'' a recommender system to impact its recommendations for \textit{all} users. 
Since YouTube's recommendation system specifically leverages ``viewing habits'' of similar users to make personalized recommendations \cite{howytrecommendations}, we surmise that the use of our intervention by a small but sizable number of users has the potential to mitigate biased recommendations for YouTube users at large through this indirect process.


Nevertheless, it is imperative that social media platforms, and YouTube in particular, make their recommendation systems more transparent to end users, scholars, and regulators \textit{and also} proactively and seriously work on mitigating the well-documented ideological biases in its recommendation algorithm. 
The dangers of these biased recommendations to citizens, groups, and society need to be factored in when prioritizing engagement and revenue over social order and also need to be accounted for by regulators and policymakers who have been slow to responding to problems in social media platforms.

Action is needed especially because our audit found more biases among right-leaning users and our intervention showed that it takes considerably more effort to debias YouTube recommendations for these users as well (see also \cite{Chen_Pacheco_Yang_Menczer_2021} for similar biases on Twitter, where right-leaning accounts did not experience exposure to moderate information and produced increasingly partisan content, partly due to bots and social networks). Although these right-wing biases are not very pronounced, they are there, detected, and visible. They are also problematic given that YouTube tends to be more popular among right-leaning versus left-leaning users \cite{munger2022right, pew2020youtubenews}. 
According to the ``supply and demand'' framework  \cite{munger2022right}, YouTube allows extremist content publishers (supply) to reach communities that are radicalized or at risk of radicalization (demand) and further incentivizes publishers to post extremist content (e.g., monetization), driving up the supply of this content, which is then demanded by these communities. 
Inasmuch as this vicious cycle occurs particularly on the right \cite{munger2022right}, and inasmuch as it is the political right that is increasingly disengaged from mainstream media and institutional politics \cite{kalmoe2019lethal} and decreasingly committed to democracy \cite{Bartels_2020}\footnote{Consider the evidence that most Republicans agreed that ``the traditional American way of life is disappearing so fast that we may have to use force to save it,'' over 40\% agreed that ``a time will come when patriotic Americans have to take the law into their own hands'' \cite{Bartels_2020}.}, the corrosive impact of biased YouTube recommendations on American -- and likely international -- democracy has to be addressed.

It is our hope that our work offers the much-needed evidence and a multidisciplinary intervention framework to foster efforts against ideological biases in social recommendation systems.

\subsection*{Acknowledgement}
This research is funded in part by the Robert N. Noyce Trust and the National Science Foundation.

\bibliographystyle{unsrt}
\bibliography{bibliography}

\end{document}